\documentclass[useAMS,usenatbib]{aa}
\usepackage{txfonts}
\usepackage{natbib}
\bibpunct{(}{)}{;}{a}{}{,}
\usepackage{aalongtable}
\usepackage[dvips]{graphicx}
\begin{document}
\title{The Luminosity Function of X-ray point sources in Centaurus A}
\author{R. Voss \and M. Gilfanov}
\institute{Max Planck Institut f\"ur Astrophysik, Germany\\
email:[voss;gilfanov]@mpa-garching.mpg.de}
\titlerunning{LF of X-ray point sources in Cen A}
\date{Received ... / Accepted ...}

\offprints{R. Voss}

\abstract{
We have studied the X-ray point source population of \object{Centaurus A} 
(NGC 5128) using data from four archival \textit{CHANDRA} observations. 
We detected 272
point-like X-ray sources within a radius of 10$\arcmin$ from the
centre. Approximately half of these are CXB sources, with the
remaining half being LMXBs. The spatial distribution of the LMXBs, both
azimuthally averaged and 2-D, is consistent with the distribution of
the \textit{K}-band light observed in the 2MASS survey. 
After correction for the incompleteness effect we constrain the LMXB
luminosity function down to $\sim 2\times 10^{36}$ erg s$^{-1}$, much
lower than previous studies of LMXBs in elliptical galaxies. 
The obtained XLF flattens significantly below $L_X\sim 5\times
10^{37}$ erg s$^{-1}$ and follows the $dN/dL\propto L^{-1}$ law in
agreement with the behaviour found earlier for LMXBs in the Milky Way
and in the bulge of M31.  

}

\maketitle

\section{introduction}

\textit{CHANDRA} observations of the bright end, $\log(L_X)\ga 37.5-38$, of
X-ray point source populations in nearby elliptical galaxies found a
rather steep luminosity distribution with a differential power law
index in the $\sim 1.8-2.5$ range \citep[e.g.][]{intro8,discussion3}.     
This is noticably steeper than X-ray luminosity function (XLF) slopes
in spiral and starburst galaxies, $\sim 1.6$ \citep{binaries5}.  
This difference reflects the difference in the composition of the
X-ray populations in the early and late type galaxies, dominated by 
low- and high-mass X-ray binaries, respectively.
Extension of the luminosity range available for the study down to
$\log(L_X)\sim 36$ revealed a much more complex shape of the XLF of
low-mass X-ray binaries (LMXBs).   
It has been shown to flatten considerably at the
faint end and to follow the $dN/dL\propto L^{-1}$ power law
below $\log(L_X)\la 37-37.5$ \citep{binaries6}. 
Motivated by observational results, \citet{discussion5} and
\citet{discussion4} suggested that the slope of the LMXB XLF in
different luminosity regimes is defined by predominantly different 
sub-types of low-mass X-ray binaries. 
In the sample of \citet{binaries6} the faint end of
the LMXB XLF was represented by the bulges of two spiral galaxies
only -- the Milky Way and M31. On the other hand, the X-ray binaries
in elliptical 
galaxies and spiral bulges could be formed by different mechanisms and
have different evolution histories and, consequently, different
luminosity distributions. 
It is therefore important to complement theoretical advances in 
understanding the XLF of X-ray binaries with firm observational
constraints on its behaviour based on a broad range of galactic 
types, especially at the low luminosity end.

Centaurus A (Cen A) is candidate for such a study.
It is massive enough to contain a sufficient number of LMXBs and, on the
other hand, is sufficiently nearby to reach luminosities below $\sim
10^{37}$ erg~s$^{-1}$ with moderate observing times. 
It has been widely studied in X-rays, and it has been observed
10 times with \textit{CHANDRA}. These observations have
been used to obtain information about the nucleus \citep{intro9}, the
interstellar medium \citep{intro6}, the jet \citep{intro4,intro7} the
shell structures \citep{intro5} and the  off-centre point
source population \citep{intro3}.
The objective of the present study is the population of LMXBs 
in Cen A, namely their spatial and luminosity
distribution. Studying the latter, we will focus specifically on the
low luminosity domain, $\log(L_X)\sim 36.5-37.5$, whose importance has
been emphasized above.  
Combining 4 observations and accurate incompleteness correction
enabled us to investigate sources with luminosity by a factor of $\sim
5-10$ lower than in previous studies. 

Cen A has a strongly warped dust disc with evidence for star formation,
and optical images show a system of filaments and shells. This
is probably due to a recent merger \citep{intro1}.
It is the nearest active galaxy and is considered to be the
prototypical Faranoff-Riley class I radio galaxy. It has a very
compact nucleus, most likely an accreting massive black hole,
with strongly varying intensity.
Emanating from this nucleus are milliarcsecond radio jets and a
subrelativistic radio/X-ray jet extend
$\sim$ 6$\arcmin$ towards NE of the nucleus. Radio lobes extending
NE and SW are seen.
An exhaustive review of Cen A can be found in \citet{intro2}.

\begin{table*}
\begin{center}
\caption{The \textit{CHANDRA} observations used in this paper.}
\label{obs}
\begin{tabular}{lcccccc}
\hline\hline
Obs-ID & Date & Instrument & Exp. Time  & R.A. & Dec. & Data Mode \\
\hline
0316 & 1999 Dec 05 & ACIS-I & 36.18 ks & 13 25 27.61 & $-$43 01 08.90 & FAINT \\
0962 & 2000 May 17 & ACIS-I & 36.97 ks & 13 25 27.61 & $-$43 01 08.90 & FAINT \\
2987 & 2002 Sep 03 & ACIS-S & 45.18 ks & 13 25 28.69 & $-$43 00 59.70 & FAINT \\
3965 & 2003 Sep 14 & ACIS-S & 50.17 ks & 13 25 28.70 & $-$43 00 59.70 & FAINT \\
\hline
\end{tabular}
\end{center}
\end{table*}

\begin{figure*}
\begin{center}
\includegraphics[clip=]{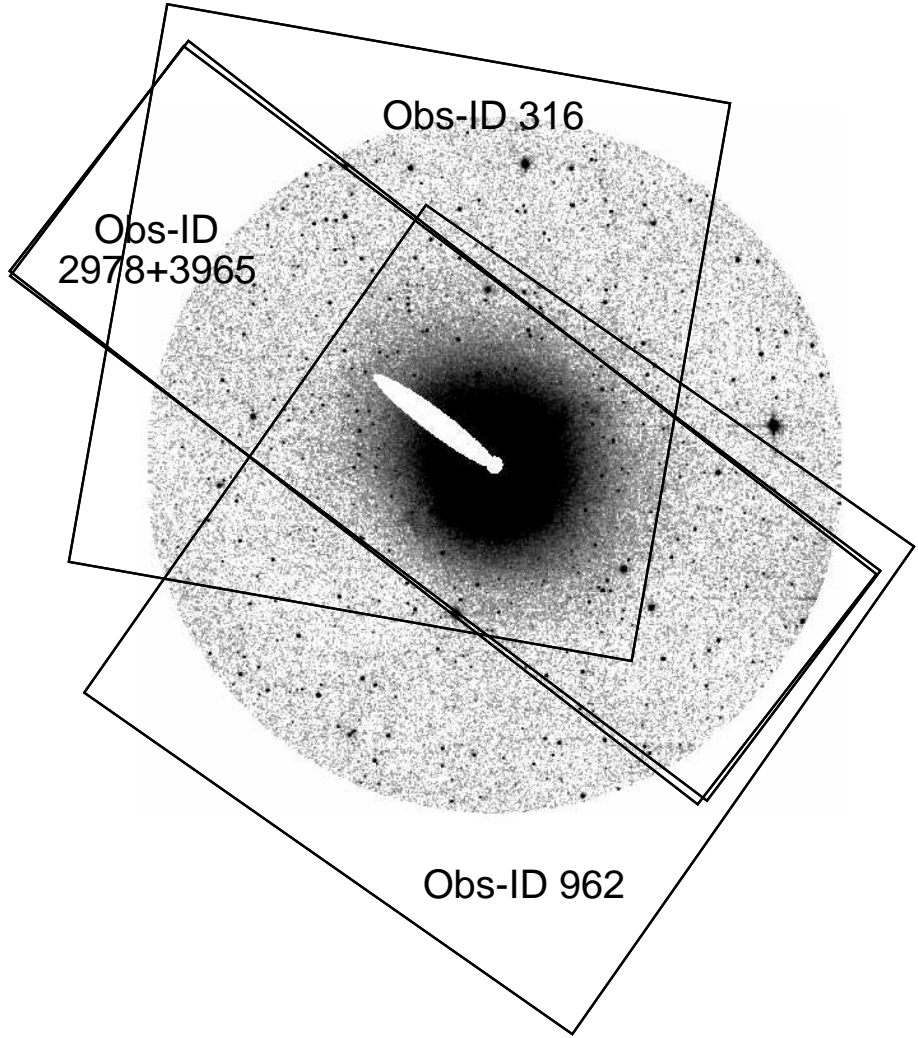}
\end{center}
\caption{The 2MASS \textit{K}-band image of the region of Cen A analysed in
this paper. The radius of the image is 10$\arcmin$. Also shown are the areas
covered by the four \textit{CHANDRA} observations.}
\label{2MASS}
\end{figure*}

The paper is structured as follows.  
In Sect. 2 we describe the data sets and the basic data
preparation and analysis. The source list cleaning procedures are
presented in Sect. 3, together with source identifications. In this
section we also deal with possible periodic variability of the most
luminous sources. 
The properties of the population of X-ray binaries, as well as the
background X-ray sources
are analysed and compared with previous studies in Sect. 4,
including the spatial distribution of the sources and their luminosity
function. Sect. 5 gives the conclusions.  We adopt a distance of 3.5 Mpc to
Cen A, and that (R.A.,Dec.) = (13~25~27.6,~-43~01~08.8) is the centre
of the galaxy.

\section{Data analysis}
\label{sec:data}

The analysis in this paper is based on four \textit{CHANDRA} observations,
two of them made with the ACIS-I array (OBS-ID 316 and 962), and the 
other two with the ACIS-S array (OBS-ID 2978 and 3965). Information
about the observations is listed in Table \ref{obs}; their fields of
view overlaid on the K-band image of the galaxy are shown in
Fig.\ref{2MASS}.

Together these four 
observations cover most of Cen A within a 10$\arcmin$ radius from
the centre.
The data preparation was done following the standard {CIAO\footnote{http://cxc.harvard.edu/ciao/} threads
(CIAO version 3.1; CALDB version 2.28), and limiting the energy range to 0.5-8.0 keV. 
The ACIS chips sometimes experience flares of enhanced background.
For point source detection and luminosity estimation it is not
necessary to filter out weak flares, since the increased exposure
time outweighs the increased background. We did not find any flares
strong enough to filter them out. 

We used CIAO wavdetect to detect sources. This program is the most widely used
for point source detection in \textit{CHANDRA} data. Some of the parameters we
have changed from the default values. Most important are the scales. 
We have used the
$\sqrt{2}$-series from 1.0 to 8.0. This gives a wide
enough range of source sizes to account for the variation
in point spread function (PSF) from the inner parts of Cen A to the 
parts 10$\arcmin$ from
the centre as well as enough middle scales. We also used
maxiter=10, iterstop=0.00001 and bkgsigthresh=0.0001.
The effect of changing these parameters is that more iterations
are done in the process of removing sources when creating
backgound files, at the expense of computing time.
Finally we set the parameter eenergy=0.8 (the encircled fraction of
source energy used for source parameter estimation), which gives 
larger areas for source parameter estimation at the risk of source 
merging, see Sect. 3.

First we detected sources in the inner region of Cen A covered by all
four observations. From these sources we then chose 40 that are bright 
enough to have the positions determined precisely and that existed in all
four observations. We used these sources to determine the average
positions of the sources and the offsets for the individual
observations. The statistical uncertainties of the source positions are
typically 0.3$-$0.5 pixel. Assuming that 
the errors are uncorrelated gives an uncertainty of $\sim$0.05 pixel
in the calculated offsets of the observations.
Using CIAO
dmtcalc we then corrected the aspect and events file for each observation.
The corrections applied are listed in Table \ref{skycor}. This step was 
performed in order to make the observations aligned for combination,
not to get better absolute astrometry, which will be dealt with in
Sect. 3.

\begin{table}
\caption{The corrections applied to the \textit{CHANDRA} aspect files
to align the observations.}
\label{skycor}
\begin{center}
\begin{tabular}{lcc}
\hline\hline
Obs-ID & Correction West & Correction North\\
\hline
0316 & $-$0.73 pixel$^{\ast}$ & $-$0.42 pixel\\
0962 & $+$1.58 pixel & $+$1.44 pixel\\
2978 & $-$0.53 pixel & $-$0.18 pixel\\
3965 & $-$0.31 pixel & $-$0.85 pixel\\
\hline
\end{tabular}
\end{center}
\footnotesize{ $^{\ast}$1 pixel is 0.492$\arcsec$}
\end{table}

We used CIAO reproject\_events to reproject
observations 316, 962 and 2978 into the sky coordinates
of observation 3965. The files were then merged and the wavdetect task
was applied again to the combined image. The output count rate for
each detected source is calculated  inside a source cell and the local
background is subtracted. For  
each source we extracted the PSF using CIAO psfextract task and
calculated the percentage of the counts expected to lie inside each
source cell. This was done for each of the four observations, and the
result was averaged using the values of the exposure maps as weights.  
For most sources this percentage is above 97 per cent, and only four
sources have values lower than 70 per cent. 
An exposure map was created for each of the observations, assuming the
energy distribution to be a powerlaw with photon index of 1.7 and
Galactic absorption of 8.4$\times 10^{20}$cm$^{-2}$
\citep{analysis6}. We assumed the same spectrum to convert the
observed count rates to unabsorbed source luminosities. 

In the very inner parts of Cen A there is strong X-ray emission
from hot gas and the central AGN. At the same time there is a
large number of point sources within a small area making crowding
a serious problem. We have therefore excluded the area within a
radius of 30 pixels ($\sim15 \arcsec$) from the centre of the
galaxy. 

Simulations using the observed source distribution as input
show that excluding this inner region limits crowding to less
than 4\% of the sources (sec. \ref{sec:verify}).
Also the part of the galaxy dominated by the X-ray jet has
been excluded. The excluded regions are evident from Fig.\ref{2MASS}.

\begin{figure*}
\begin{center}
\includegraphics{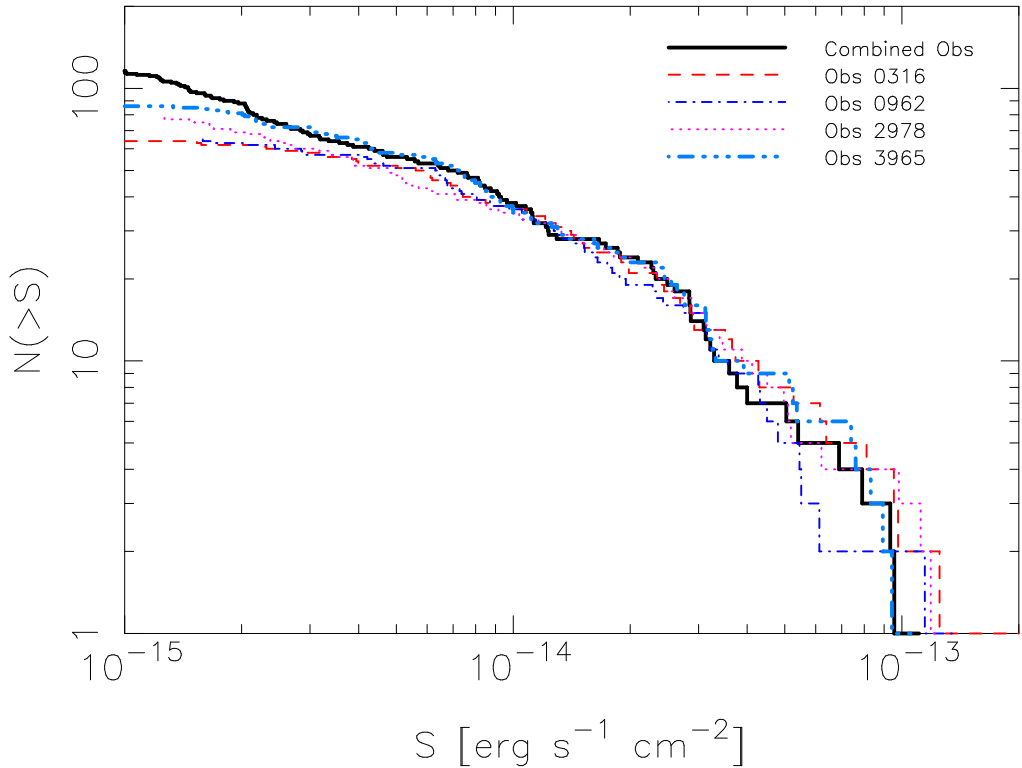}
\end{center}
\caption{Comparison of the cumulative $\log(N)-\log(S)$ distribution found 
in each of the separate observations and in the combined image. Only
sources from the inner region, contained in all four observations,
have been used and incompleteness correction has not been applied.
The incompleteness begins to have effect at a few $\times 10^{-15}$
erg s$^{-1}$cm$^{-2}$ for the combined image and at 
$\sim 7-8 \times 10^{-15}$erg s$^{-1}$cm$^{-2}$ for the individual
observations.} 
\label{compare}
\end{figure*}

In each of the four observations, readout streaks caused by the bright
central region of Cen A are seen. As in different observations they
cover different regions of the image, for each streak we have searched
for sources and estimated their parameters using a combined image of
the observations, excluding the one containing this streak.

To check for differences between the four observations and
between the individual observations and the combined observation, we
have created the cumulative point source luminosity function for each,
taking only sources from the central region, where all four observations
overlap, see Fig. \ref{2MASS}. The results can be found in Fig. 
\ref{compare}. A large
fraction of the sources are variable. For these sources, using the
luminosities estimated from the combined image is equal to using the  
average luminosities. We used the Kolmogorov--Smirnov (KS)-test to
compare the $\log(N)-\log(S)$ distribution obtained in the individual 
observations with that based on the combined data.
To minimize incompleteness effects only sources with fluxes
higher than 3$\cdot 10^{-15}$ erg s$^{-1}$ cm$^{-2}$ were used. The
lowest probability found was 68 per cent (for Obs-ID 316). This
confirms that the source variability does not modify the flux
distribution of the point sources in a galaxy like Cen A at a
detectable level.

\section{The source list}

Several effects can compromise the source list generated from
CIAO wavdetect. This includes extended sources and false sources due to
background fluctuations. The background due to the diffuse emission
is high, especially in the inner parts of Cen A, and many
structures can be seen in the image.  The ``bubble'' $\sim$5$\arcmin$ 
south-west of  the centre is an example
\citep{intro6}. Some of these 
structures might be misinterpreted as point sources. We have visually
inspected the images and for each source compared the photon
distribution with the distribution expected from the PSF. As a result
we rejected 18 sources. As indicated by the
shapes, none of the rejected sources is likely to be a
supernova remnant. 
Some of the rejected sources are filamentary
structures in the diffuse component and the rest are caused by local
variations in the emission of the diffuse component. The
characteristic length scale of the latter is $\gtrsim$100 pc. 
Due to the low luminosity 
of the rejected sources, it is not possible to classify them 
according to their spectra.

Another potential problem could be merging of sources. We have used a high 
value (80 per cent) of the enclosed percentage of PSF in CIAO wavdetect 
because it gives a good estimation of source parameters. On the other
hand, such a high value in some cases leads to two sources being
detected as one source. To check for this, we ran CIAO wavdetect again
with smaller enclosed percentages of the PSF. We find no sources that 
are merged because of the high enclosed percentage of the PSF. 

After the filtering, the final list of X-ray sources contains 272
objects. It is presented in Table \ref{tab:list}. \citet{intro3} analysed the two
ACIS-I observations of Cen A (Table \ref{obs}) and detected 246
X-ray sources. Of these, 205 sources are located within $r<10\arcmin$
of the center of the galaxy analysed here. 184 of these sources are
in our source list, which therefore contains 90 previously undetected
sources. The $\sim 1/3$ increase
in the total number of detected sources is due to a factor of 
$\ga 2-4$ increase in the exposure time of the main body of the
galaxy (Fig.\ref{2MASS}, Table \ref{obs}).

\subsection{Background and foreground sources}

A fraction of the detected sources are foreground or background
objects. Some (but not all) of them can be identified using either 
their X-ray spectra, or from observations at other wavelengths. 
Since this paper concerns the statistical properties of
the X-ray point source population, we have adopted the following
strategy. We exclude foreground sources as much as possible (6 such sources
are excluded, see Sect. \ref{sect:optical}), but do
not attempt to remove background sources, which are by far the most
significantly contaminating factor (about half of the detected sources
are background sources, see Sect. \ref{normCXB}). Their contribution to the surface
brightness and luminosity distributions is instead taken
into account in the statistical sense, based on the results of the 
cosmic X-ray background (CXB) source counts.

\subsection{Optical identifications}\label{sect:optical}

We check the absolute astrometry using USNO-B1.0 \citep{USNO} and 
GSC 2.2 \citep{GSC} catalogues. 
We find that for a search radius of 2.0$\arcsec$ the rms deviation of the
positions is 1.1$\arcsec$. This is comparable to the
quoted  positional uncertainties of the optical catalogues as well as
that of the Chandra X-ray source list, confirming reasonable astrometric
accuracy of the latter. Adding a systematic shift of 0.5$\arcsec$ in any
direction results in larger rms deviations. The number of matches 
is significantly higher  
than the expected number of chance coincidences. For the search radius  
of 2.0$\arcsec$ the expectation value is $\sim 8$ with 37 matches found
for USNO-B1.0 and $\sim 3$ with 18 matches found for GSC 2.2. 
 
For the actual identification of Chandra sources we used the results
of the dedicated optical studies of the Cen A region by \citet{clean2},
\citet{clean1} and \citet{WHH}. 
Although the former three surveys were aimed specifically at globular
cluster population of Cen A they also have identified a number of
foreground stars, H$_\alpha$ emittors and several AGNs. We also used
results of \citet{clean4}. In total we identified 6 X-ray sources as
foreground stars, leaving 266 sources of presumably extragalactic
origin -- either intrinsic Cen A sources or background AGNs. 
Of these, 37 were identified with the globular clusters in Cen A.  
The results of this work are presented in
column 8 of Table \ref{tab:list}. 

About $\sim2/3$ of the USNO and GSC matches were found to
be globular clusters or likely globular clusters in \citet{clean2} and 
\citet{clean1}. 
The remaining 12 out of 37 sources do not appear in these 
papers. This is close to but slightly higher than the number of 8 
random matches expected for the value of the search radius used in the 
analysis. Some of these sources also might be background AGNs or
undetected globular clusters. Therefore we kept them all in the
sample. We note that excluding them from the following analysis does
not change our results in any significant way.

\subsection{H$\alpha$-sources}

\begin{figure}
\includegraphics{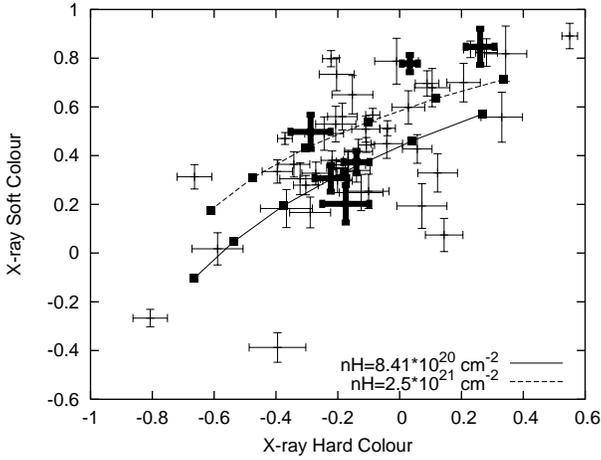}
\caption{The color--color diagram of the brightest, $>200$ counts,
sources within 5 armin from the centre of Cen A. The sources
coinciding with H$_\alpha$-emitting regions are shown in  bold. 
For reference, the two lines show the hardness ratios of power law
spectra for two different values of absorption.   
The filled squares are at photon indices of 0.5, 1.0, 1.5,
2.0, 2.5, 3.0 from right to left. 
The hard and soft colours are defined as HC=(H-M)/(H+M),
SC=(M-S)/(M+S), where S,M and H are the number of photons detected in
the 0.5-1.0 keV, 1.0-2.0 keV and 2.0-8.0 keV energy range
respectively.} 
\label{halpha}
\end{figure}

Eight sources within 4 $\arcmin$ from the centre of Cen A coincide with
H$\alpha$-emitting regions found in \citet{clean1}. 
All of them are located in the dust lanes region, have X-ray
luminosities in the $10^{36}-5\cdot 10^{37}$ erg~s$^{-1}$ range  and can
potentially be associated with high-mass X-ray binaries. 
The optical magnitudes of the  H$\alpha$ sources indicate that they may be 
young star clusters as well as individual X-ray binaries.
In order to search for further indications of the high-mass X-ray binary
(HMXB) nature of these
sources we have compared their spectral properties with other sources 
and searched for periodic variability in their X-ray emission.  
As discussed in more detail below, no coherent
pulsations were detected from any of the bright X-ray sources,
although the upper limits are at a rather moderate level of $\sim
25$ per cent pulsed fraction.

The accreting X-ray pulsars, constituting the vast majority of the 
neutron star HMXBs, are known to have notably harder spectra in the
$\sim$ 1--20 keV energy range  than LMXBs and often show significant
intrinsic absorption. Therefore comparison of the spectral properties
of the H$\alpha$ objects with other X-ray sources (which in the
central part of Cen A are mostly LMXBs, Sect. \ref{sec:xlf}) can
help to clarify the nature of the former.
However, the X-ray colour-colour diagram of the sources within 
5$\arcmin$ from the centre of Cen A, shown in Fig \ref{halpha},
does not reveal systematic differences between  H$\alpha$ and other
sources, nor have we found any systematic differences from the direct
spectral fits of the brightest sources.

Comparing Fig \ref{halpha} with Fig. 4 of \citet{prestwich} and noting
the slight difference in energy bands, it can be seen that the main
part of our sources is located in the region corresponding to LMXBs.
There is a small population of harder sources, of which two are H$\alpha$
objects and also a few softer sources. From their position in the diagram,
they could be HMXBs and thermal supernova remnants, respectively.
Such identifications are not possible with to the colours alone for two
reasons. One is that the absorption inside Cen A varies strongly with
position, which has the effect of enhancing the scatter of LMXBs in
the diagram. The second reason is that there is a contribution of
CXB sources. This population is known to consist of two subpopulations,
a hard and a soft one. These populations would be expected to coincide
with the HMXBs and the supernova remnants, respectively, in the diagram.

As our results were not conclusive enough, we decided to keep the
H$\alpha$  sources in the sample, bearing in mind that their nature
still needs to be clarified. Due to their relatively small
number they do not significantly affect the following analysis of the
spatial and luminosity distributions.

\subsection{Globular cluster sources}

37 X-ray sources coincide with known globular clusters. 
Interpreting this number, one should  take into account that only
$\sim 20-25$ per cent of the expected number of globular clusters in
Cen A have been identified \citep{WHH}. The identified sample is
strongly biased, both 
with respect to the spatial distribution of the clusters and their
luminosity distribution. Furthermore, the detection of the globular
clusters is not independent of the X-ray observations, as X-ray
source catalogues have been used to search for globular clusters
\citep[e.g.][]{clean1}. It is therefore not possible to perform a rigorous
comparison of the luminosity function and spatial distribution of the  
globular cluster X-ray sources with the sources residing outside
globular clusters. 
Considering the sources brighter than 
$3\cdot 10^{37}$ erg~s$^{-1}$ (i.e. unaffected by incompleteness
effects) there are 15 known globular cluster X-ray sources, whereas
the number of sources outside globular clusters is 40. If the expected 
number of 22 CXB sources (see Sect. \ref{normCXB})
is subtracted, we find 18 'field' LMXBs outside (or in undetected)
globular clusters. Above this luminosity the XLF of globular cluster
sources is similar to that of the field LMXBs. Below this luminosity 
the field LMXB XLF is much steeper than that of globular cluster
sources. This can easily be caused by incompleteness effects,
which cannot be correcetd for without knowledge of the spatial
distribution of the globular cluster sources (see Appendix).
We also note that in the outer region there are 7 globular cluster
sources,  with L$_X\ge 10^{37}$ erg~s$^{-1}$ whereas we expect $\sim$9
LMXBs in total (see Sect. \ref{LMXB}).

\subsection{Search for coherent pulsations}

We searched for periodic variability in the light curves of the sources with
more than 400 detected source counts and more luminous than 6.0$\cdot 10^{37}$
erg s$^{-1}$ (24 sources in total). Each \textit{CHANDRA} observation was 
tested
separately.  Events were extracted from the 4 sigma source ellipses in  
{\tt wrecon} and the light curves with $\approx 3.2$ sec time
resolution were produced. The power spectra were calculated using the
STARLINK\footnote{http://www.starlink.ac.uk} task {\tt period}. 
Pulsations were searched for
in the range of trial periods from $P\approx 6.4$ s, defined by the
Nyquist frequency of the Chandra time series, to $P=2000$ s. 
Except 3 sources showing variability due to the telescope
dithering carrying them over the detector edge, in only one did the
power exceeded the level corresponding to 99 per cent confidence. 
The period of 55.8 s was found for the source \#135 (Table
\ref{tab:list}) in Obs 2978 and had a significance of 99.4
per cent. This significance takes into account the number of trial
periods in one power spectrum but not the number of power spectra
analyzed (74). In the other 3 observations of the source the power
density spectrum did not show any signs of pulsations at this period. 
Given the total number of power spectra investigated
it is likely that this detection is a result of a statistical
fluctuation. 
Even for the most luminous sources, pulsed
fractions of $\sim$25 per cent would be needed for detection at the
99 per cent confidence level.

\section{Populations of X-ray sources in the field of Centaurus A}
\label{sec:xlf}

In the central $r<10\arcmin$ of Cen A (excluding the nucleus and the
jet, Fig.\ref{2MASS}, Sect. \ref{sec:data}) we detected 136 sources with
$L_X>10^{37}$ erg~s$^{-1}$ and  252  ($\approx 321$ after the incompleteness
correction) sources with $L_X>2\cdot 10^{36}$ erg s$^{-1}$ (Table
\ref{tab:list}, \ref{tab:numbers}).

\begin{table}
\renewcommand{\arraystretch}{1.1}
\centering
\caption{Expected and observed numbers of point sources 
(section \ref{sec:numbers}.}
\begin{tabular}{crrrrr}
\hline\hline
$L_X$ & \multicolumn{4}{c}{Predicted} & Obs.\\
erg~s$^{-1}$ & LMXB & HMXB & CXB$^{(1)}$ & Total$^{(1)}$ & Total$^{(2)}$\\
\hline
$>10^{37}$& 81 & 10 & 34 (47) &125 (138) &136 \\
$>2\cdot 10^{36}$& 155 & 27 & 98 (135) & 280 (317) & 321\\
\hline
\end{tabular}\\
\parbox{\columnwidth}{
(1) -- the CXB numbers are based on the soft (hard) band counts from
\citet{analysis3}, see Sect. \ref{CXB};
(2) -- after the incompleteness correction
}
\label{tab:numbers}
\end{table}

\subsection{Expected numbers}
\label{sec:numbers}

\subsubsection{Low mass X-ray binaries}\label{LMXB}

LMXBs are related to the population of old stars, and there
is therefore a correlation between their number  and the
stellar mass of a galaxy \citep{binaries6}. 
In order to estimate the expected number and luminosity distribuition
of LMXBs we used a \textit{K}-band image from the 2MASS Large Galaxy Atlas
\citep{2MASS} and integrated the flux emitted in the parts of Cen A
analysed in this paper. This gives the \textit{K}-band luminosity  
of $L_{K}=8.6\cdot 10^{10}~ L_{\odot}$. To convert it to the
stellar mass we use the color-dependent \textit{K}-band mass-to-light ratio
from \citet{binaries7}. For the extinction corrected optical
color of Cen A, $(B-V)\approx 0.88$, the mass-to-light
ratio is $M_{\ast}/L_K\approx 0.76$.
This gives the stellar mass of $5.5\cdot 10^{10}$ $M_{\odot}$, assuming
that the absolute \textit{K}-band magnitude of the sun is equal to
$M_{K,\odot}=3.39$.
Using the results of \citet{binaries6} we predict $\approx$81 LMXBs
with $L_X>10^{37}$ erg s$^{-1}$, and $\approx$155 with $L_X>2\cdot 10^{36}$
erg s$^{-1}$.

\subsubsection{High mass X-ray binaries}
\label{HMXB}

Being young objects, HMXBs are associated with star formation and, as 
expected for an elliptical galaxy, are by far a less significant
contribution to the population of X-ray binaries than LMXBs. 
In terms of absolute rates,  star formation in Cen A is mostly
associated with the dust disk. From their analysis of IRAS data,
\citet{binaries2} found the total far infra-red (FIR) luminosity of
the Cen A disc to be $9.7\cdot 10^9$ $L_{\odot}$ 
($L_{\odot}=3.8\cdot 10^{33}$ erg s$^{-1}$).
From this luminosity we subtracted the emission from the central
region which  is mostly due to the active nucleus, $1.5\cdot
10^9~L_{\odot}$, and corrected the distance from the 5 Mpc
assumed in \citet{binaries2} to the 3.5 Mpc adopted in this
paper. This gives $L_{\rm FIR}\approx4.0\cdot 10^9$ $L_{\odot}$. 
Assuming that the total infrared luminosity is
$L_{\rm TIR}\approx 2 L_{FIR}$ and using the SFR calibration of
\citet{binaries4} we find ${\rm SFR}\approx 1.4$ M$_{\odot}$
yr$^{-1}$.  We used the calibration of \citet{binaries5} to calculate the
expected number of HMXBs  
(see comment in \citealt{shty} regarding
the normalization). 
From this we get the expectation of $\approx$10 HMXBs
brighter than $10^{37}$ erg s$^{-1}$, and $\approx$27 sources brighter
than $2\cdot 10^{36}$ erg s$^{-1}$.

\subsubsection{Background X-ray sources}
\label{CXB}

To estimate the number of background sources we use the
results of the CXB $\log(N)-\log(S)$ determination by
\citet{analysis3}. We use the source counts in the soft and
hard bands (their Eq. 2) and  convert the fluxes to the
0.5--8.0 keV band, assuming a powerlaw spectrum with a photon index of
1.4. For the total area of our survey of 0.079 deg$^2$  we obtain from the source counts in the soft band 
$\approx$34 CXB sources above the flux corresponding to 
$10^{37}$ erg s$^{-1}$, and $\approx$98 above $10^{36}$ erg
s$^{-1}$. From the hard band counts the predicted numbers are 
$\approx$47 and $\approx$135 sources. 
The predictions based on the soft and hard $\log(N)-\log(S)$ differ
because of the well recognized fact that source counts in different
energy bands and flux regimes are dominated by different types of
sources. This is further discussed in Sect. \ref{normCXB}.
Furthermore the normalization of the CXB source counts is subject
to uncertainy due
to the cosmic variance. Its rms amplitude is $\sim 20-25$\% 
\citep[see e.g.][]{analysis5}.

\bigskip
The results of the above calculations are summarized in
Table \ref{tab:numbers}. For the total number of point sources, the
agreement between observed and predicted values is surprisingly good, given
the amplitude of uncertainties involved. 
In the following two subsections we derive from 
the data, and compare with the predictions, the abundances of
individual types of X-ray sources. This is done in two independent
ways -- based on the radial distribution of the sources (Sect.
\ref{spatial}) and on their flux/luminosity distribution (Sect.
\ref{normCXB}).

\begin{figure}
\begin{center}
\includegraphics[clip=]{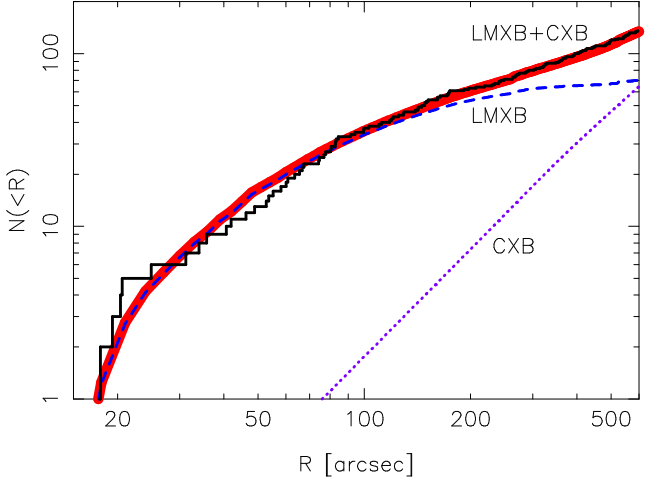}
\end{center}
\caption{The radial distribution of observed sources (solid line),
compared to the best fit model (thick grey line) and the contributions
of LMXBs and CXBs} 
\label{fig:radial}
\end{figure}

\subsection{Spatial distribution of point sources}
\label{spatial}

We begin with the azimuthally averaged radial profile 
(Fig.\ref{fig:radial}). 
As it follows from the results of the previous section the two major 
contributiors to the population of point sources in the field of Cen A 
are low-mass X-ray binaries ($\sim$ 1/2--2/3 of the sources, depending
on the luminosity)  and background AGNs ($\sim$ 1/3--1/2). 
Correspondingly, we model the observed distributions as a
superposition of  two functions, describing their respective
contributions. 
The spatial distribution of the LMXBs has been shown \citep{binaries6} 
to follow, to first approximation, the distribution of the stellar
mass. The latter can be represented by the distribution of the \textit{K}-band
light and was computed using the \textit{K}-band image of Cen A from the 2MASS
Large Galaxy Atlas \citep{2MASS}. The density of the CXB sources can
be assumed to be flat on the angular scales under consideration, therefore
the CXB growth curve is proportional to the enclosed solid angle.  
In computing both radial profiles we took into account
that some areas were excluded from the analysis (Fig.\ref{2MASS}). 
The only free parameter of the model is the ratio of normalizations of
the LMXB and CXB distributions.
The (unknown) distribution of HMXBs has not been included as it is
unlikely to exceed 10\% of the total number of sources 
(Sect. \ref{HMXB}). 

The model  has been compared with the observed
distribution of sources more luminous than $10^{37}$erg s$^{-1}$.
This value of the luminosity threshold was chosen in order to include
as many sources as possible and, on the other hand, to keep
incompleteness effects insignificant. The model adequately
describes the data (Fig.\ref{fig:radial}) as confirmed with the
KS-test, with a probability of 96 per 
cent. The best fit LMXB fraction, determined from the Maximum Likelihood (M-L)
fit to the the unbinned radial distribution data, is  $51.7\pm 5.9$
per cent, corresponding to $70.3\pm 10.0$ LMXBs and $65.7\pm 9.8$ CXB
sources.   Compared to Table \ref{tab:numbers}, the abundance
of LMXBs is  surprisingly close to the expectated value. The number of
CXB sources, on the other hand, is higher than the expectation. This
will be further discussed in Sect. \ref{normCXB}.

The same LMXB+CXB model was also compared with the radially averaged
azimuthal and two--dimensional distributions. 
The KS test of the unbinned two--dimensional
distribution of the point sources \citep[e.g.][]{radial1} gave a
probability of $24\%$. The  azimuthal distribution of the
sources within 5$\arcmin$ (to exclude the outer regions dominated by
CXB) has the KS-probability of $10-20$ per cent depending on the
starting point. 
Also we checked whether there was any azimuthal dependence on
the radial profiles, by dividing the observations into two and
three slices and comparing them using the KS-test. Trying a lot
of different angles, we found no evidence for such a dependence.
Due to the low number of sources, such evidence would not be
found unless the effect was strong.

This analysis confirms that within the statistical accuracy of the
data, the spatial distribution of the LMXBs is consistent with that of
the K-band light. This implies, in particular, that no additional
component corresponding to HMXBs is required by the data. However,
this result is not very constraining, given the rather small expected
number of HMXBs,  $\approx 10$.

\subsubsection{Sensitivity of the spatial distribution analysis}

In order to probe the sensitivity of the above analysis we performed
the following test. The LMXB distribution was streched with
respect to the center of the galaxy by some scale factor, the new best
fit value of the CXB to LMXB ratio was found using same method as
before, and the consistency of new best fit model with the data was
checked with the KS test. Then the range of values or the scale
factor was found beyond which the KS probability decreased below 
$5$ per cent indicating deteriorated quality of the approximation.  
The following ranges for the scale factor values were obtained: 
$0.4\la\eta \la1.9$ for the radial profile analysis 
and $0.2\la\eta \la2$ for the 2-dimensional image.

These numbers indicate a rather moderate sensitivity of the
spatial distribution analysis. Sensitivity limitations of 
this kind are unavoidable when analysing individual galaxies.
Further exposure of the inner 10$\arcmin$ of the galaxy can improve
the luminosity limit below which incompleteness effects have to be 
taken into account. Observations with the telescope pointing to
the outskirts of the galaxy could be useful too, as they could
help to constrain the local CXB normalization. Also a
very careful study of the source distribution at luminosities
where incompleteness is a problem could increase the sensitivity.
Another approach is to study combined source density distributions for
several (many) galaxies.

\subsection{Source counts and the cosmic X-ray background source density}
\label{normCXB}

We divided Cen A into three annuli according to the ratio of
predicted numbers of LMXBs and CXB sources: $r<2.5\arcmin$, $r=2.5-5
\arcmin$, $r=5-10 \arcmin$). 
The inner and outer regions are expected to be dominated by LMXBs and
CXB sources respectively, while the middle one contains comparable
numbers of sources of both types  (e.g. Fig.\ref{fig:radial}).
In analysing the luminosity functions and $\log(N)-\log(S)$
distributions we used the procedure described in Appendix \ref{app} to
correct for incompleteness effects.

\begin{figure}
\begin{center}
\includegraphics[width=\columnwidth ,clip=]{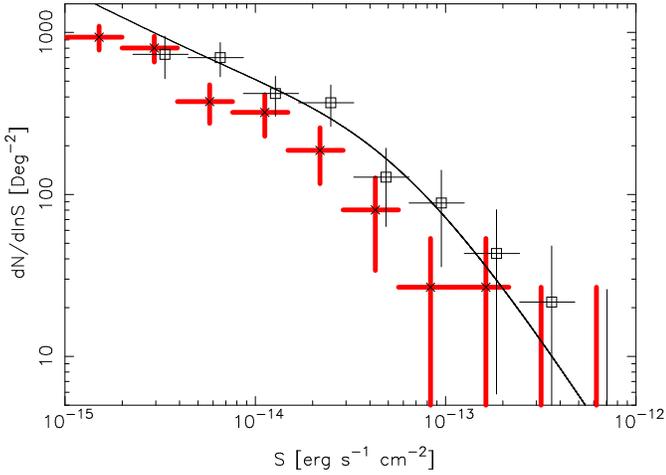}
\end{center}
\caption{The source counts (open squares) in the outer region 
($5\arcmin<r<10\arcmin$). The predicted contribution of LMXBs is
subtracted. The thick solid line shows the CXB $\log(N)-\log(S)$ from 
\citet{analysis3}.
with best fit normalization from this paper.
For comparison the source counts in the CDF-N Obs-ID 1671 are shown (crosses).}
\label{difouter}
\end{figure}

We estimate the normalization of the CXB $\log(N)-\log(S)$ distribution
from the source counts in the outer region. This region is far enough
from the inner parts of the Cen A to keep the number of sources
related to the galaxy low, while close enough to the aimpoints of the 
observations to have a reasonable sensitivity. 
In this region the incompleteness corrected number of sources with 
the 0.5--8 keV  flux exceeding $2.7\cdot$10$^{-15}$ erg s$^{-1}$
cm$^{-2}$ (luminosity $4.0\cdot 10^{36}$ erg s$^{-1}$) is 101.3 from 
which 13.4 
are expected to be LMXBs. The implied number of CXB sources is
$\approx 88$, which we compare with the results of the radial profile
analysis from the previous section and with results of dedicated CXB
source counts. For this comparison we express the CXB normalization in
units of the number of sources per deg$^{2}$ with 0.5--8.0 keV flux
$S_X >6.8\cdot 10^{-15}$ erg s$^{-1}$ cm$^{-2}$. The results of the
CXB surveys are transformed to the 0.5--8.0 keV energy range assuming
a power law spectrum with the photon index of 1.4. We used the source
counts in both their soft (0.5--2.0 keV) and hard (2.0--8.0 keV)
bands. The results are shown in the two columns in the upper part of
Table \ref{tab:cxb}. 
The last two lines in Table \ref{tab:cxb} present the results of the
radial profile analysis and of the source counts in the 
$r=5\arcmin-10\arcmin$ ring. Note that although these two numbers are
not statistically independent, they are obtained from  
different considerations. The radial profile analysis is based on  
sources with $L_X\ge10^{37}$ erg~s$^{-1}$ in the entire $r\le 10 \arcmin$
region and relies heavily on the assumption about the spatial
distribution of the LMXB component. The source counts in the outer
region use all sources in the $5\arcmin\le r\le 10\arcmin$ with a
luminosity above
$10^{36}$ erg~s$^{-1}$ and are significantly less dependent on the assumption
of the LMXB spatial distribution.

\begin{table}
\caption{CXB normalization found in various surveys and in this paper.} 
\label{tab:cxb}
\centering
\begin{tabular}{lcc}
\hline\hline
Survey & soft band & hard band  \\
\hline
CDF-S & 332$\pm$70 & 686$\pm$71\\ 
CDF-N & 437$\pm$80 & 791$\pm$73\\
Cappelluti et al., 2004 & 350$\pm$28 & 419$\pm$43\\
Cowie et al., 2002 & $-$ & 456$\pm$30\\
Moretti et al., 2003 & 422 & 579\\
\hline
Obs--ID 1671 (CDF-N)& \multicolumn{2}{c}{$519\pm 71$}\\
Radial profile (Cen A)& \multicolumn{2}{c}{$832\pm 124$}\\
$5\arcmin<r<10\arcmin$ counts (Cen A)& \multicolumn{2}{c}{$804\pm 86$}\\
\hline
\end{tabular}\medskip

\parbox{\columnwidth}{
The normalization is expressed as the number of sources per deg$^{2}$ 
with 0.5--8.0 keV flux $S_X >6.8\cdot 10^{-15}$ erg s$^{-1}$
cm$^{-2}$. The two columns give the numbers computed from the soft and
hard band counts respectively.
The for CDF-fields data listed in the upper part of the table are from
\citet{analysis1}, in the lower part -- from this paper.  
}
\end{table}

\begin{figure}
\begin{center}
\includegraphics[width=\columnwidth ,clip=]{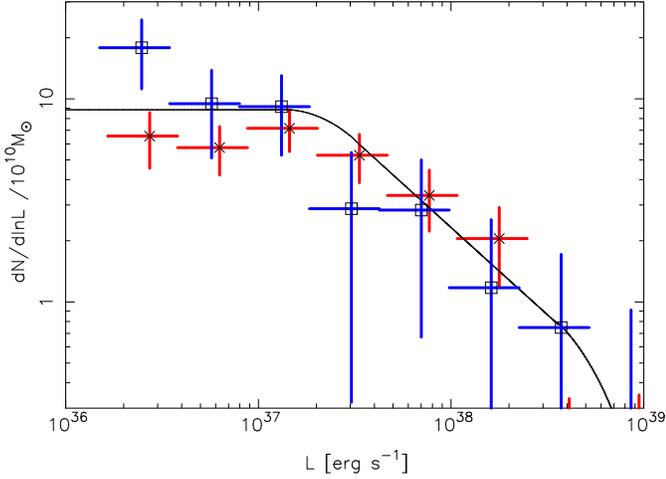}
\end{center}
\caption{The differential luminosity functions of LMXBs in the innermost
(stars) and middle (squares) annuli, normalized to 10$^{10}$ $M_{\odot}$ of
stellar mass.  
The CXB contribution is subtracted using the results of Sect.
\ref{normCXB}. The solid line shows the average LMXB XLF from
\citet{binaries6} smoothed with the boxcar filter with the
logarithmically constant width equal to the bin width in the observed
XLFs.} 
\label{fig:xlf_lmxb}
\end{figure}

As it has been already mentioned, there is a significant difference
between the normalizations found from the hard and soft
bands. This is related to the fact that different types of sources give
dominant contributions to the hard and soft bands. Theoretically, the
two bands can be reconciled using different spectral shapes for the
the flux conversion, but this would introduce additional uncertainties
and an investigation of this kind is beyond the scope of this paper. 
In addition, there is
also a considerable spread in the CXB normalizations in the same
energy band obtained in different surveys. This spread is partly due
to the cosmic variance 
and partly it is likely to be caused by the difference in the analysis
procedures and relative calibrations of different instruments.

In order to do a direct comparison with the empty fields source counts
in the $0.5-8.0$ keV energy band, we have
analysed one observation from the CDF-N (Obs-ID 1671), using the
same data analysis procedure as we used for Cen A. The column density
of neutral hydrogen was set to 1.5$\cdot 10^{20}$ 
cm$^{-2}$ \citep{analysis6}. 
To avoid incompleteness effects, we only used sources observed in
regions with exposure above $4.4\cdot 10^7$~s~cm$^2$. This limits the field
to 0.058 deg$^2$. Above a flux of $2.7\cdot 10^{-15}$ erg s$^{-1}$ cm$^{-2}$
(equal to the flux used to extimate the CXB normalization in the outer
region of Cen A) we find 53 sources. This number can be directly
compared with 88 CXB sources detected in the outer annulus in the Cen
A field (by chance the two areas coincide).
In order to facilitate comparison with the other CXB surveys, we
transform this number to the units of Table \ref{tab:cxb},
using the $\log(N)-\log(S)$ from the soft band of \citet{analysis3}.    

Even with the spread in values found from the various surveys, 
the CXB normalization in the Cen A field appears to be higher than the
typical numbers obtained in the dedicated CXB studies, with the exception
of the hard band counts of the Chandra Deep Fields, according to the
analysis of \citet{analysis1}. The  latter two excluded, the density of
CXB sources appears to be enhanced by a factor of $\sim
1.4-2$. Although this is larger than the rms variation between
different fields typically quoted in the literature, $\sim 20-25$ per
cent, the observed number is not exceptionally high and still lies
within the spread of the CXB  density values
\citep[e.g.][]{analysis5}.

\begin{figure*}
\begin{center}
\includegraphics[clip=]{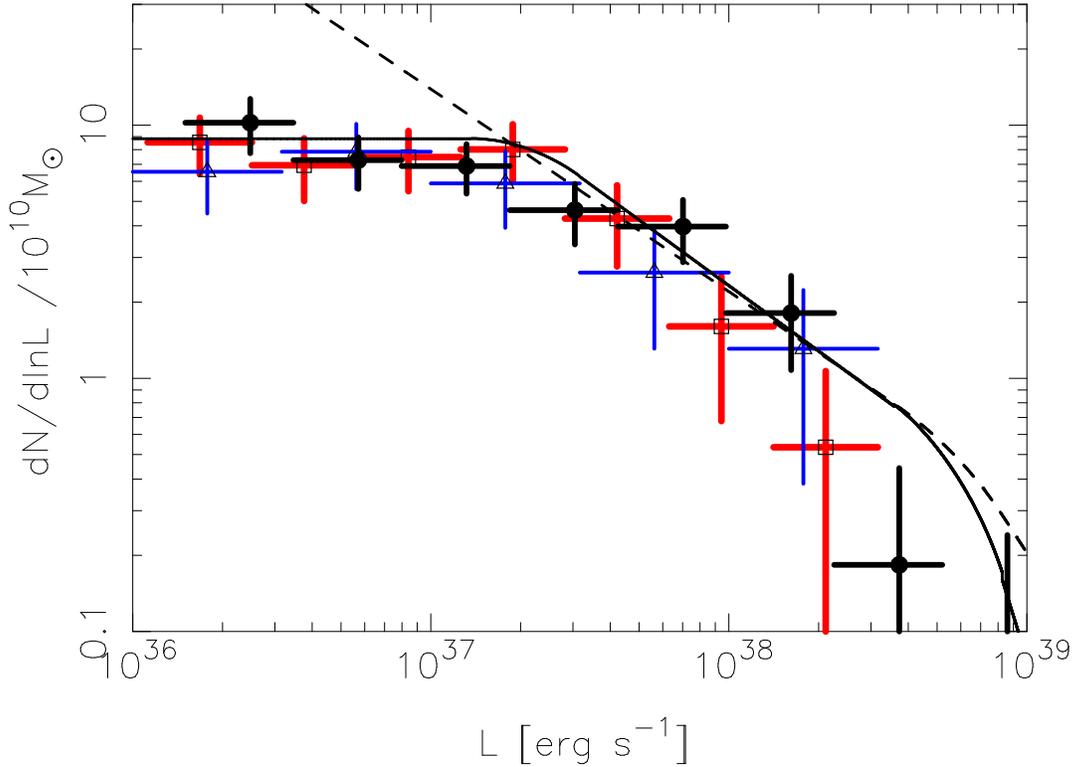}
\end{center}
\caption{The luminosity function of LMXBs in the inner $r\le 5\arcmin$
of Cen A (the CXB contribution subtracted) in comparison with LMXB
XLFs in the Milky Way (triangles) and M31 (squares) 
\citep[from][]{binaries6}. The latter two are multiplied by constant
factors of  1.7 and 0.6 respectively. The solid line shows the average
LMXB XLF in the nearby galaxies as determined by \citet{binaries6},
the same as in Fig.\ref{fig:xlf_lmxb}.  The dashed line shows the
average LMXB XLF from \citet{discussion3} and its extrapolation
towards low luminosities. Its normalization was chosen to
approximately match our observations. 
}
\label{fig:xlf_lmxb2}
\end{figure*}

\subsection{LMXB X-ray luminosity function}
\label{LMXBsec}

The luminosity function of LMXBs determined from the two inner regions
is shown in Fig.\ref{fig:xlf_lmxb}. In subtracting the contribution of
CXB sources we used the $\log(N)-\log(S)$ distribution from
\citet{analysis3}  with the normalization determined in the
Sect. \ref{normCXB}. While 
the CXB contribution is unimportant in the innermost region 
$r\le 2\farcm 5$, it accounts for about half of the sources
in the middle region $2\farcm5 \le r \le 5\arcmin$. 
As is obvious from Fig.\ref{fig:xlf_lmxb}, both distributions are
consistent with each other and with the average LMXB XLF in the local 
galaxies determined by \citet{binaries6}, with the possible exception of
the the lowest luminosity bin of the middle region, which deviates by
$\sim 1.5\sigma$.

To further constrain the parameters of LMXB XLF in Cen A we fit the
luminosity distribution in the inner region with a power law with two
breaks, identical to the one used in \citet{binaries6}. 
Since there are no sources luminous enough to constrain the
upper break and the slope beyond that, we have fixed them at the
average values: $L_{b2}=5.0\cdot 10^{38}$ erg s$^{-1}$, 
$\alpha_3=4.8$. 
The best fit values of  other parameters are:
the low luminosity slope $\alpha_1=1.02^{+0.12}_{-0.13}$, a break at
$L_{b1}=5.0^{+1.0}_{-0.7}\cdot 10^{37}$ erg s$^{-1}$ and 
a slope after the break $\alpha_2=2.6\pm 0.4$.
The slopes refer to the differential distribution, the parameter
errors are $1\sigma$ statistical errors only. Notice that the
break value found for differential XLFs is systematically higher
than the break value found for cumulative XLFs, using the same
data, see e.g. \citet{kaaret}.
These parameters are insensitive to whether the CXB component is
accounted for or not.

A large uniformly analysed sample of the XLF of LMXBs in elliptical
galaxies was presented by \citet{discussion3}. They find an average
differential slope of $1.8\pm0.2$ in the luminosity range
$L_X=$ a few $\times 10^{37}$ to $5\times 10^{38}$ erg s$^{-1}$.
This is consistent with our results from the inner region. A
KS-test gives 73 per cent probability that the observed luminosity
distribution above $L_X=1.0\cdot 10^{37}$ erg s$^{-1}$ could be produced 
by their LMXB XLF. On the other
hand it is clear that at the faint end of the XLF the extrapolation of
their results is  inconsistent with our observations. For
sources more luminous than  $L_X=5.0\cdot 10^{36}$ erg s$^{-1}$, a
similar KS-test gives 3.4 per cent, and for lower luminosities the
probability decreases further.

The LMXB XLF based on the combined data of $r\le 5\arcmin$ is plotted
in Fig.\ref{fig:xlf_lmxb2} along with luminosity distributions of
LMXBs in the Milky Way and M31. This plot further illustrates the
qualitative and quantitative similarity of the LMXB luminosity
distributions in Cen A and bulges of spiral galaxies.
This is the first study to extend the LMXB XLF in elliptical galaxies
below $\sim$few$\times 10^{37}$ erg~s$^{-1}$.
Spiral and elliptical galaxies have different evolutionary histories
and it could differ in the properties of their LMXB populations. 
As demonstrated here, the luminosity functions nevertheless seem very
similar, except for the break luminosity which could be somewhat
higher in Cen A than in  the Milky Way and M31. Whether this reflects
a systematic difference between LMXBs in galaxies of different type is
yet to be investigated.

\subsection{$X/M_\ast$ ratios}

In the inner (middle) region there are 53 (27) sources with 
$L_X>10^{37}$ erg~s$^{-1}$, with an integrated luminosity of
$L_X=2.3\cdot 10^{39}$ ($1.3\cdot 10^{39}$)  erg~s$^{-1}$. We expect
the CXB contribution to be $N_{\rm CXB}$=3.8 (13.4), corresponding  
to a luminosity of $1.9\cdot 10^{38}$ ($6.7\cdot 10^{38}$) 
erg~s$^{-1}$. From the \textit{K}-band light we estimate that the
stellar mass is 
3.6$\cdot 10^{10}$ (1.3$\cdot 10^{10}$) $M_\odot$, and this gives
us the ratios $N_X$/$M_\ast$=13.7$\pm$1.9 (10.5$\pm$2.0) sources per
$10^{10}$ $M_\odot$ and $L_X$/$M_\ast$=6.4 (4.8) $\times 10^{38}$
erg~s$^{-1}$ per $10^{10}$ $M_\odot$. The values for the two regions
are consistent. 
They are also in a good agreement with the values for different
nearby galaxies listed in Table 2 of \citet{binaries6} as well as with
the average values of $\left<N_X/M_\ast\right>=14.3$ and
$\left<L_X/M_\ast\right>=8.0\cdot 10^{38}$ erg/s per $10^{10}$
$M_\odot$. 

The $X/M_\ast$ ratios obtained in this paper are by a factor 
of 2 lower than the values  for Cen A in \citet{binaries6}. 
He reported problems in approximating the
multi-aperture K-band photometry data for Cen A galaxy. Indeed, we
recomputed the K-band luminosity for the same region using the 2MASS
K-band image and obtained  $\approx 2$ times larger number. 
This explains the lower values of $N_X$/$M_\ast$ and $L_X$/$M_\ast$
found in this paper. As these numbers are derived from the real K-band
images rather than from extrapolation of the multi-aperture K-band
photometry, they better represent the true values of the 
$X/M_\ast$ ratios.

\section{Summary and conclusions}

We have used archival data of \textit{CHANDRA} observations to
study statistical properties of the point source population of Cen
A. Our primary goal was to investigate the faint end of the LMXB
luminosity distribution in an elliptical galaxy and to compare it with
LMXB XLF in bulges of spiral galaxies. 

To achieve this we assembled as deep a survey of the central
part of the galaxy as permitted by the  available data and implemented
an adequate correction for the incompleteness effects.

Cen A is the closest giant
elliptical galaxy and the only one with enough exposure time by
\textit{CHANDRA} to perform such a study. As Cen A is a merger remnant,
the stellar and LMXB population might differ from those of less
disturbed giant ellipticals. It is therefore important to further
perform deep studies of the X-ray source population of more normal
early-type galaxies.

Using a combined image of four ACIS observations  (Table \ref{obs},
Fig.\ref{2MASS}) with the total
exposure time of 170 ks we have  detected 272
point-like sources within 10$\arcmin$ of the nucleus of Cen A. 
The luminosity of the weakest detected source is $\approx 9\cdot 10^{35}$
erg~s$^{-1}$ (assuming a distance of 3.5 Mpc), while the source sample
starts to be affected by the incompleteness effects below $\sim
10^{37}$ erg~s$^{-1}$ (Fig.\ref{correction}). 
After correction for incompleteness, the total number
of sources with $L_X\ge 2\cdot 10^{36}$ erg~s$^{-1}$ is $\approx 321$. 
This number is in good agreement with the prediction based on
the stellar mass, the star formation rate in Cen A and the density
of CXB sources (Table \ref{tab:numbers}).  About
half of the detected sources are expected to be X-ray binaries in Cen
A, mostly LMXBs; the vast majority of the remaining sources are
background galaxies constituting the resolved part of the CXB.   

The spatial distribution of the detected sources can be well
described by a sum of two components. Of these, one has a density
proportional to the \textit{K}-band light (Fig.\ref{fig:radial}) and the
other is uniform accross the Cen A field.
We interpret this as that the former represents low-mass X-ray 
binaries in Cen A while the latter accounts for the resolved part of
the CXB.  
The normalization of the LMXB component agrees well with the average
value derived for the local galaxies by \citet{binaries6}.
The normalization of the uniform component and source counts in the
exteriors of the galaxy appear to indicate an overabundance
of the CXB sources in the direction of Cen A by a factor of 
$\sim 1.5$ or, possibly, more (Table \ref{tab:cxb}, Fig.\ref{difouter}).

After applying the incompleteness correction and subtracting the
contribution of CXB sources we were able to recover the the LMXB
luminosity function   in the inner $r\le 5\arcmin$  down to 
$L_X\sim 2\cdot 10^{36}$ erg~s$^{-1}$
(Fig.\ref{fig:xlf_lmxb},\ref{fig:xlf_lmxb2}). This is by a factor of
$\sim 5-10$  better than achieved previously for any
elliptical galaxy \citep{intro3,discussion3}.  
The shape of the luminosity distribution is consistent with the
average LMXB XLF in nearby galaxies derived by \citet{binaries6}
and for the bright end by \citet{discussion3}. In particular, we demonstrate 
that the LMXB
XLF in Cen A flattens at the faint end and is inconsistent with
extrapolation of the steep power law with differential slope of
$\approx 1.8-1.9$ observed above $\log(L_X)\sim 37.5-38$ in the 
previous studies of elliptical galaxies. Rather, the LMXB XLF in Cen A
has a break at  $L_X\approx (5\pm1)\cdot 10^{37}$ erg~s$^{-1}$ below which
it follows the   $dN/dL\propto L^{-1\pm0.1}$ law, similar to the 
behaviour found in the bulges of spiral galaxies.

\begin{acknowledgements}
This reaserch has made use of \textit{CHANDRA} archival data provided by
the \textit{CHANDRA} X-ray Center and data from the 2MASS Large Galaxy
Atlas provided by NASA/IPAC infrared science archive. We would like to
thank the referee for useful comments.
\end{acknowledgements}

\appendix

\section{Correction for incompleteness}
\label{app}

\begin{figure}
\begin{center}
\includegraphics[width=\columnwidth ,clip=]{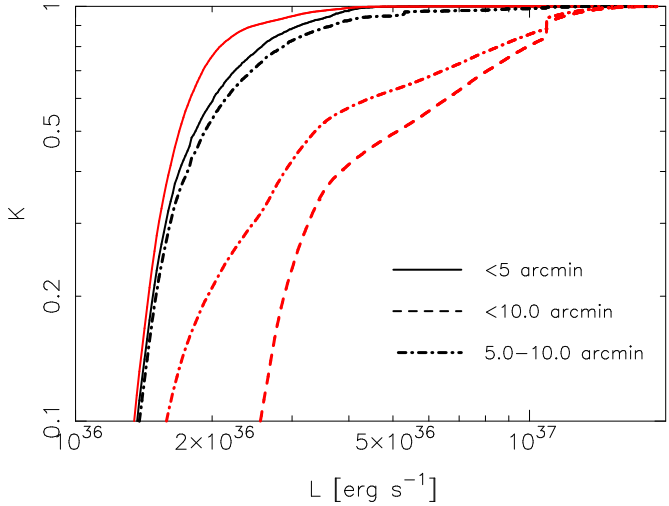}
\end{center}
\caption{The sample incompleteness as a function of the luminosity for
the inner ($<5 \arcmin$), outer (5--10 $\arcmin$) and full region.
The black lines are calculated using the source density proportional
to the \textit{K}-band light distribution.
The grey lines are calculated assuming a uniform source density.
}
\label{correction}
\end{figure}

The variations of the diffuse background level and deterioration of
the point spread function at large off-axis angles lead to  variations
of the point-source sensitivity accross the Chandra images.
In the case where several observations with different pointing
directions are combined, this effect is further amplified by
the non uniform exposure of the combined image. As a result, the
completness of the source sample at the faint end is compromised. 
A trivial solution to this problem is to define a conservative
sensitivity limit, which is high enough to be achieved everywhere
across the image. Although simple in implementation, this method has
a disadvantage that  a noticeable fraction of the 
source has to be thrown away. Nevertheless, it has been used, with few 
exceptions \citep[e.g.][]{discussion3, shty}, in the majority of the
earlier  studies of the point source populations in galaxies.  
A more effective approach to the problem is to define the correction
function to the flux/luminosity distribution, which accounts for the
sensitivity variations accross the image. 
For a uniform distribution of sources this correction function simply
accounts for the dependence of the survey area upon the energy flux or
count rate. This is the case, for example, in the CXB studies.
In a more complex case of a non-uniform distribution of point
sources, the observed and real flux distributions are related via:   
\begin{equation} 
\left(\frac{dN}{dS}\right)_{\rm obs}=\int_{S_0\le S} 
\frac{dN}{dS}\, \Sigma(x,y)\, dx dy
\label{eq:incompl1}
\end{equation} 
where $\Sigma(x,y)$ is the surface density distribution of  point
sources, and for given flux $S$ the integration is performed over the
part of the image where the local sensitivity $S_0(x,y)$ satisfies the
condition $S_0(x,y)\le S$. If the flux distribution does not depend on
the position, it can be easily recovered from the above equation. 
Importantly, knowledge of the spatial distribution of sources is 
required in order to recover the flux distribution and vice versa.
If both flux and density distributions are unknown, the sample
incompleteness can not be properly accounted for.  
The problem is further complicated by the contribution of the CXB
sources, having a diferent spatial and flux distributions:
\begin{eqnarray} 
\left(\frac{dN}{dS}\right)_{\rm obs}=\int_{S_0\le S} 
\left(\frac{dN}{dS}\right)_{\rm LMXB}\, 
\Sigma_{\rm xrb}(x,y)\, dx dy
\nonumber
\\
+\int_{S_0\le S} 
\left(\frac{dN}{dS}\right)_{\rm CXB}\, 
\Sigma_{\rm cxb}(x,y)\, dx dy
\label{eq:incompl2}
\end{eqnarray} 

For the practical implementation of the correction procedure, knowledge
of the source detection algorithm is of course required.
The  {\tt wavdetect} task \citep{ap10} correlates the image with  a
Mexican  Hat function and registers sources with the correlation value 
exceeding a threshold value. The latter is estimated
numerically based  on the user-specified threshold significance.
For each of the used detection scales we computed the threshold
sensitivity on a grid of the positions on the image (16 azimuthal
angles, 40 radii from the centre of Cen A). At each image position the
PSF was obtained from the CALDB PSF library for each of the four
individual observations and then combined with the exposure times as
weights. The local background levels were found from the normalized
background maps created by {\tt wavdetect}. The sensitivity for any
given position on the image was found from interpolation of the grid
values. The sample incompleteness is described by the incompleteness
function: 
\begin{equation}
K(L)=\sum_{L_0(i,j)\le L}\Sigma(i,j)
\end{equation}
where $i,j$ are the pixel coordinates and $L_0(i,j)$ is the 
position-dependent sensitivity. 
Depending on the desired normalization
of the $K(L)$, the the density distribution $\Sigma(i,j)$ can be
normalized to unity or, for example, be given in the units of $M_\odot$
per pixel of the image.

\begin{figure}
\begin{center}
\includegraphics[width=\columnwidth ,clip=]{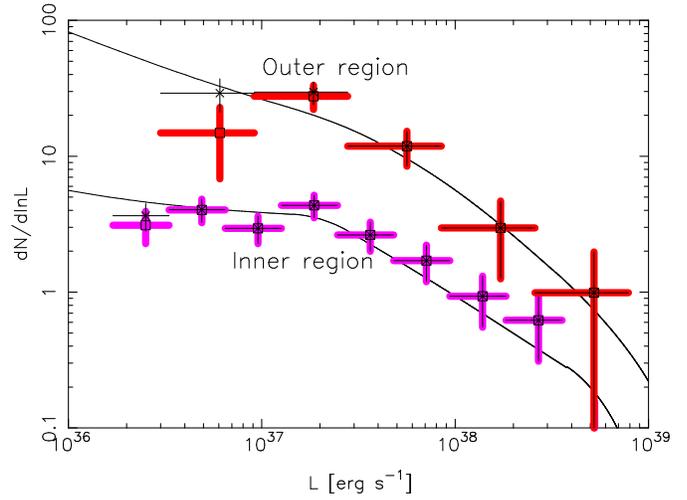}
\end{center}
\caption{Simulated luminosity functions for the inner, $r\le 5\arcmin$,
and outer, $r\ge 5\arcmin$, regions. The input distributions (solid
lines) are compared with the results obtained from the analysis of the
images, done in the same way as the analysis of the real data. Both
data corrected (asterisks) for incompleteness and 
uncorrected (squares) are shown. The normalization of the inner
region has been divided by 10 for clarity. }
\label{simulations}
\end{figure}

If the CXB contribution can be neglected (eq.\ref{eq:incompl1}), 
the corrected luminosity distribution can be obtained giving the
weight $1/K(L)$ to a source of luminosity $L$. For the ML fits the
model should be multiplied by the $K(L)$.  

For the general case of eq.\ref{eq:incompl2}, the incompleteness
function  $K(L)$ should be calculated for the CXB and
LMXB components separately. For the LMXBs density distribution  we
used the \textit{K}-band image, the CXB distribution was
assumed to be uniform. The corresponding incompleteness functions
are shown in Fig. \ref{correction}. They demonstrate clearly 
importance of the spatial distribution of sources.

The results of one of our simulations are shown in
Fig.\ref{simulations}. In these simulations the background map from
the {\tt wavdetect} and the 
\textit{K}-band image were azimutahlly averaged. The flux/luminosity
distributions for the CXB sources and LMXBs were assumed in the form
described in Sects. \ref{normCXB} and \ref{LMXBsec}. 
In the simulations, the sources were randomly drawn from the assumed
spatial and luminosity distributions and projected to the image using
the PSF data from CALDB. The image of expectation values, containing
the diffuse component and the point source contribution was then
randomized assuming Poission statistics. The final image was analysed
using the same chain of tasks as applied to the real images.
The simulated and obtained luminosity distributions are shown in
Fig.\ref{simulations}. 

The Figs.\ref{correction} and \ref{simulations} demonstrate that given
the pattern of Chandra observations of Cen A, incompleteness effects 
are not of primary concern in the inner $r\la 5\arcmin$ region at the
luminosoties above $\sim (2-3)\cdot 10^{36}$ erg/s. 
They should be taken into account, however, for the source counts in
the entire image  and in the outer ring.

\subsection{Verification of the incompleteness correction}
\label{sec:verify}

\begin{figure}
\begin{center}
\includegraphics[width=\columnwidth ,clip=]{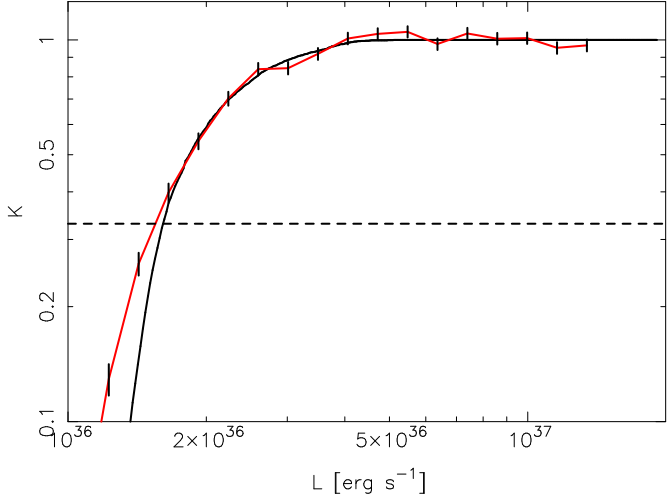}
\end{center}
\caption{The incompleteness correction function for LMXBs within 5\arcmin from the
centre of Cen A. The solid line is the function used throughout this paper, whereas
the red line is found using the backward correction method of \citet{KF}. The dashed
line marks the correction limit. Sources with a correction larger than this are not
included in the analyses carried out in this paper.}
\label{fig:check}
\end{figure}

Although simulations described above show that the correction
procedure is adequate for the analysis of Cen A, their accuracy is
limited by the Poisonian statistics. 
Such accuracy limitations are intrinsic to full
simulations of individual galaxies because  the number of sources thaimit 
can be put into a simulation is limited by the crowding effect. 
Another disadvantage of these simulation is that we used smoothed
background maps produced by {\tt wavdetect}, which, in addition, were
azimuthally averaged.

In order to perform a more accurate and sensitive check of our
incompleteness correction procedure we used a more direct, but also
more computationally expensive, method, similar to the backward
correction method suggested by \citet{KF}. In this method sources are
placed one at a time on the real (unprocessed) observed image. For
each simulated source the source detection and photometry are
performed with the {\tt wavdetect} task and then the original image  
is restored. As each source is  put on the original undisturbed image,
an arbitrarily large number of sources can be simulated.
The incompleteness function is given by the source detection
efficiency and can be computed as a ratio of the flux distribution of
the detected  sources to the input flux distribution.

Using this method we checked the incompleteness correction for the
region within 5\arcmin  ~from the centre of Cen A, for the LMXB component. As
above, the source distribution was assumed to follow the
\textit{K}-band light, and the differential luminosity function was
chosen to follow 1/L. The sources were put on the image using the
same method as in the simulations above, utilizing the CALDB PSF
library together with the exposure map of the observations. To reduce
statistical errors, we simulated 20,000 sources. 

The resulting incompleteness correction, together with the correction 
function utilized throughout this paper, is shown in figure
\ref{fig:check}. As it can be seen from the figure, the two curves
differ at low fluxes, corresponding to $L_X\la 1.5\cdot 10^{36}$
erg/s. The difference is caused by two effects. 
Firstly, the Eddington bias is neglected in our incompleteness
correction calculations whereas it is automatically included in the
simulations. The second reason is that in the {\tt wavdetect} task 
the source detection and countrate calculation are based on two
different calculations. The source detection uses the correlation
values of the wavelet transform to determine the source significance,
whereas the photometry is performed on the original image.  
Although there is a linear relation between the expectation values of
the source countrate and wavelet correlation, the measured numbers 
are subject to statistical fluctuations. This is ignored 
in the correction procedure, which uses  the wavelet
correlation values to both calculate the source significance
and source flux.
This effect is only important at low numbers of counts, where only a few
sources are detected. These weak sources are excluded from our
luminosity function analysis as we include only sources with the
detection efficiency of $\ge 1/3$. 
For the weakest source used to plot 
Figs.\ref{difouter}--\ref{fig:xlf_lmxb2}
the difference between two correction factors is 4.6\%.
In Fig. \ref{fig:xlf_lmxb2}, for example, the lowest luminosity bin
would decrease by $\sim$ 2\%.
This accuracy of the incompleteness correction is sufficient  
for the analysis presented in this paper.

{\footnotesize
\begin{longtable}{rcrccrrrccl}
\caption{The list of point like X-ray sources within $r<10\arcmin$
from the center of Cen A.}\\
\hline\hline
ID & CXO Name & Dist  & R.A.       &  Dec.     & Total Cts & Source Cts & Error & Luminosity  &
Type & ID reference\\
(1) & (2) & (3) & (4) & (5) & (6) & (7) & (8) & (9) & (10) & (11)\\
\hline
\endfirsthead
\caption{continued.}\\
\hline\hline
ID & CXO Name & Dist  & R.A.       &  Dec.     & Total Cts & Source Cts & Error & Luminosity  &
Type & ID reference\\
(1) & (2) & (3) & (4) & (5) & (6) & (7) & (8) & (9) & (10) & (11)\\
\hline
\endhead
\hline
\endfoot
1 & CXOU J132526.9-430052 & 17.5 & 13 25 27.0 & -43 00 52.8 & 499 & 266.8 & 22.3 & $2.36\cdot 10^{37}$ & \\
2 & CXOU J132529.1-430114 & 18.1 & 13 25 29.2 & -43 01 14.6 & 219 & 78.8 & 14.8 & $7.17\cdot 10^{36}$ & H$_\alpha$ & mrfa-45\\
3 & CXOU J132526.4-430054 & 19.1 & 13 25 26.5 & -43 00 54.6 & 1905 & 1480.7 & 43.6 & $1.28\cdot 10^{38}$ & \\
4 & CXOU J132526.7-430126 & 19.7 & 13 25 26.8 & -43 01 26.1 & 523 & 267.1 & 22.9 & $2.29\cdot 10^{37}$ & \\
5 & CXOU J132527.5-430128 & 19.8 & 13 25 27.5 & -43 01 28.5 & 991 & 705.3 & 31.5 & $5.99\cdot 10^{37}$ & GC & mrfa-053\\
6 & CXOU J132529.4-430108 & 20.2 & 13 25 29.5 & -43 01 08.3 & 498 & 297.4 & 22.3 & $2.52\cdot 10^{37}$ & GC & mrfa-044\\
7 & CXOU J132526.6-430129 & 23.3 & 13 25 26.6 & -43 01 29.4 & 118 & 51.6 & 10.9 & $5.25\cdot 10^{36}$ & \\
8 & CXOU J132525.7-430056 & 23.8 & 13 25 25.8 & -43 00 56.1 & 1862 & 1578.1 & 43.2 & $1.35\cdot 10^{38}$ & GC & mrfa-017\\
9 & CXOU J132525.2-430114 & 26.1 & 13 25 25.3 & -43 01 14.3 & 105 & 61.7 & 10.2 & $6.03\cdot 10^{36}$ & \\
10 & CXOU J132529.6-430122 & 26.4 & 13 25 29.7 & -43 01 22.5 & 68 & 31.3 & 8.2 & $3.09\cdot 10^{36}$ & \\
11 & CXOU J132525.5-430124 & 27.3 & 13 25 25.6 & -43 01 24.2 & 259 & 110.3 & 16.1 & $9.66\cdot 10^{36}$ & \\
12 & CXOU J132528.4-430137 & 30.6 & 13 25 28.5 & -43 01 37.9 & 96 & 42.9 & 9.8 & $4.01\cdot 10^{36}$ & \\
13 & CXOU J132525.1-430127 & 32.3 & 13 25 25.2 & -43 01 27.1 & 741 & 565.7 & 27.2 & $4.90\cdot 10^{37}$ & GC & mrfa-057\\
14 & CXOU J132524.7-430125 & 34.9 & 13 25 24.8 & -43 01 25.0 & 263 & 181.6 & 16.2 & $1.66\cdot 10^{37}$ & \\
15 & CXOU J132524.3-430110 & 35.8 & 13 25 24.4 & -43 01 10.4 & 199 & 121.7 & 14.1 & $1.09\cdot 10^{37}$ & \\
16 & CXOU J132527.0-430030 & 38.7 & 13 25 27 & -43 00 30.6 & 172 & 128.5 & 13.1 & $1.15\cdot 10^{37}$ & \\
17 & CXOU J132530.8-430128 & 40.5 & 13 25 30.9 & -43 01 28.4 & 67 & 37.4 & 8.2 & $3.54\cdot 10^{36}$ & \\
18 & CXOU J132523.9-430059 & 41.5 & 13 25 23.9 & -43 00 59.0 & 312 & 227.5 & 17.7 & $2.00\cdot 10^{37}$ & \\
19 & CXOU J132531.4-430057 & 43.9 & 13 25 31.5 & -43 00 57.2 & 38 & 21.6 & 6.2 & $2.05\cdot 10^{36}$ & \\
20 & CXOU J132523.8-430127 & 45.2 & 13 25 23.9 & -43 01 27.2 & 61 & 31.1 & 7.8 & $3.24\cdot 10^{36}$ & \\
21 & CXOU J132524.4-430141 & 47.6 & 13 25 24.4 & -43 01 41.2 & 252 & 168.2 & 15.9 & $1.51\cdot 10^{37}$ & H$_\alpha$ & mrfa-60\\
22 & CXOU J132527.0-430159 & 50.9 & 13 25 27.1 & -43 01 59.4 & 825 & 675.6 & 28.7 & $5.76\cdot 10^{37}$ & H$_\alpha$ & mrfa-54\\
23 & CXOU J132527.2-430016 & 52.9 & 13 25 27.3 & -43 00 16.0 & 35 & 21.2 & 5.9 & $2.23\cdot 10^{36}$ & \\
24 & CXOU J132524.1-430145 & 53.0 & 13 25 24.2 & -43 01 45.7 & 84 & 49.3 & 9.2 & $4.74\cdot 10^{36}$ & \\
25 & CXOU J132523.3-430043 & 53.1 & 13 25 23.4 & -43 00 43.6 & 79 & 48.2 & 8.9 & $4.61\cdot 10^{36}$ & \\
26 & CXOU J132523.5-430138 & 53.9 & 13 25 23.5 & -43 01 38.7 & 714 & 589.2 & 26.7 & $5.15\cdot 10^{37}$ & \\
27 & CXOU J132522.9-430125 & 54.2 & 13 25 22.9 & -43 01 25.1 & 1947 & 1780 & 44.1 & $1.53\cdot 10^{38}$ & \\
28 & CXOU J132523.0-430134 & 56.1 & 13 25 23.1 & -43 01 34.9 & 226 & 157.3 & 15 & $1.46\cdot 10^{37}$ & \\
29 & CXOU J132532.4-430134 & 58.9 & 13 25 32.5 & -43 01 34.3 & 869 & 768.3 & 29.5 & $6.41\cdot 10^{37}$ & \\
30 & CXOU J132522.3-430122 & 59.1 & 13 25 22.4 & -43 01 22.3 & 63 & 33.8 & 7.9 & $3.37\cdot 10^{36}$ & \\
31 & CXOU J132533.0-430108 & 59.9 & 13 25 33.1 & -43 01 08.0 & 264 & 213.4 & 16.2 & $1.80\cdot 10^{37}$ & \\
32 & CXOU J132523.0-430145 & 61.9 & 13 25 23.1 & -43 01 45.8 & 406 & 312.7 & 20.2 & $2.77\cdot 10^{37}$ & \\
33 & CXOU J132526.9-430004 & 64.4 & 13 25 26.9 & -43 00 04.9 & 25 & 18.5 & 5.0 & $4.84\cdot 10^{36}$ & \\
34 & CXOU J132522.1-430132 & 65.0 & 13 25 22.1 & -43 01 32.3 & 85 & 59.1 & 9.2 & $6.00\cdot 10^{36}$ & \\
35 & CXOU J132533.3-430053 & 65.0 & 13 25 33.4 & -43 00 53.1 & 507 & 427.8 & 22.5 & $3.64\cdot 10^{37}$ & H$_\alpha$ & mrfa-06\\
36 & CXOU J132525.5-430210 & 65.5 & 13 25 25.6 & -43 02 10.5 & 63 & 46 & 7.9 & $4.54\cdot 10^{36}$ & GC & mrfa-055\\
37 & CXOU J132527.4-430214 & 65.5 & 13 25 27.5 & -43 02 14.3 & 1609 & 1496.2 & 40.1 & $1.27\cdot 10^{38}$ & \\
38 & CXOU J132531.3-430203 & 68.2 & 13 25 31.4 & -43 02 03.3 & 22 & 18.8 & 4.7 & $1.74\cdot 10^{37}$ & \\
39 & CXOU J132527.6-430218 & 69.4 & 13 25 27.7 & -43 02 18.2 & 288 & 227.8 & 17 & $1.98\cdot 10^{37}$ & FS & mrfa-51\\
40 & CXOU J132521.3-430046 & 72.1 & 13 25 21.4 & -43 00 46.0 & 51 & 36.9 & 7.1 & $3.72\cdot 10^{36}$ & \\
41 & CXOU J132522.8-430017 & 72.9 & 13 25 22.9 & -43 00 17.6 & 363 & 331.7 & 19.1 & $3.02\cdot 10^{37}$ & \\
42 & CXOU J132523.7-430009 & 73.0 & 13 25 23.7 & -43 00 09.7 & 1835 & 1730 & 42.8 & $1.50\cdot 10^{38}$ & H$_\alpha$ & mrfa-21\\
43 & CXOU J132524.7-430002 & 73.2 & 13 25 24.8 & -43 00 02.6 & 116 & 85.4 & 10.8 & $7.40\cdot 10^{36}$ & \\
44 & CXOU J132524.2-425959 & 78.5 & 13 25 24.2 & -42 59 59.7 & 573 & 521.5 & 23.9 & $4.55\cdot 10^{37}$ & H$_\alpha$ & mrfa-19\\
45 & CXOU J132525.3-430223 & 78.8 & 13 25 25.3 & -43 02 23.4 & 213 & 180.1 & 14.6 & $1.63\cdot 10^{37}$ & \\
46 & CXOU J132521.7-430154 & 78.8 & 13 25 21.7 & -43 01 54.2 & 191 & 137.5 & 13.8 & $1.22\cdot 10^{37}$ & \\
47 & CXOU J132531.6-430003 & 78.9 & 13 25 31.6 & -43 00 03.3 & 1023 & 962.6 & 32 & $8.08\cdot 10^{37}$ & GC & pff-gc-210\\
48 & CXOU J132528.7-425948 & 81.2 & 13 25 28.8 & -42 59 48.6 & 1107 & 1034.7 & 33.3 & $8.67\cdot 10^{37}$ & \\
49 & CXOU J132521.2-430154 & 83.9 & 13 25 21.2 & -43 01 55.0 & 212 & 189 & 14.6 & $1.78\cdot 10^{37}$ & \\
50 & CXOU J132523.5-430220 & 84.4 & 13 25 23.6 & -43 02 20.8 & 465 & 421 & 21.6 & $3.73\cdot 10^{37}$ & \\
51 & CXOU J132521.2-430158 & 85.5 & 13 25 21.3 & -43 01 58.9 & 223 & 174.4 & 14.9 & $1.54\cdot 10^{37}$ & \\
52 & CXOU J132521.5-430213 & 93.0 & 13 25 21.6 & -43 02 13.8 & 146 & 124.7 & 12.1 & $1.15\cdot 10^{37}$ & \\
53 & CXOU J132520.8-430010 & 94.3 & 13 25 20.8 & -43 00 10.8 & 27 & 20.3 & 5.2 & $2.11\cdot 10^{36}$ & \\
54 & CXOU J132532.0-430231 & 95.8 & 13 25 32.0 & -43 02 31.5 & 535 & 490.7 & 23.1 & $4.17\cdot 10^{37}$ & H$_\alpha$ & mrfa-40\\
55 & CXOU J132525.8-425933 & 97.4 & 13 25 25.8 & -42 59 33.4 & 33 & 17.4 & 5.7 & $1.61\cdot 10^{36}$ & \\
56 & CXOU J132530.3-425935 & 98.1 & 13 25 30.3 & -42 59 35.2 & 136 & 89.8 & 11.7 & $7.58\cdot 10^{36}$ & GC & pff-gc-209\\
57 & CXOU J132518.9-430136 & 98.7 & 13 25 19.0 & -43 01 37.0 & 67 & 48.6 & 8.2 & $4.81\cdot 10^{36}$ & \\
58 & CXOU J132529.0-425931 & 98.9 & 13 25 29.0 & -42 59 31.1 & 95 & 59.8 & 9.7 & $5.10\cdot 10^{36}$ & \\
59 & CXOU J132536.6-430057 & 99.5 & 13 25 36.6 & -43 00 57.6 & 303 & 233.9 & 17.4 & $1.94\cdot 10^{37}$ & \\
60 & CXOU J132518.5-430116 & 99.8 & 13 25 18.5 & -43 01 16.3 & 578 & 523.1 & 24 & $4.58\cdot 10^{37}$ & GC & mrfa-074\\
61 & CXOU J132519.9-430203 & 100.4 & 13 25 19.9 & -43 02 03.3 & 38 & 25 & 6.2 & $2.69\cdot 10^{36}$ & \\
62 & CXOU J132518.7-430141 & 102.7 & 13 25 18.7 & -43 01 41.2 & 96 & 72.3 & 9.8 & $7.04\cdot 10^{36}$ & \\
63 & CXOU J132519.2-430158 & 104.6 & 13 25 19.2 & -43 01 58.4 & 61 & 45.1 & 7.8 & $5.07\cdot 10^{36}$ & \\
64 & CXOU J132528.2-430253 & 105.0 & 13 25 28.2 & -43 02 53.6 & 368 & 340.1 & 19.2 & $2.98\cdot 10^{37}$ & \\
65 & CXOU J132537.4-430131 & 110.2 & 13 25 37.4 & -43 01 31.8 & 109 & 81.5 & 10.4 & $6.81\cdot 10^{36}$ & GC & mrfa-033\\
66 & CXOU J132526.7-430300 & 112.0 & 13 25 26.8 & -43 03 00.4 & 188 & 159.4 & 13.7 & $1.40\cdot 10^{37}$ & \\
67 & CXOU J132518.7-430205 & 113.2 & 13 25 18.7 & -43 02 05.8 & 41 & 27.3 & 6.4 & $2.88\cdot 10^{36}$ & \\
68 & CXOU J132522.2-430245 & 113.7 & 13 25 22.2 & -43 02 45.8 & 126 & 116.9 & 11.2 & $1.09\cdot 10^{37}$ & GC & pff-gc-121\\
69 & CXOU J132520.6-425942 & 115.5 & 13 25 20.7 & -42 59 42.1 & 22 & 12.7 & 4.7 & $1.30\cdot 10^{36}$ & \\
70 & CXOU J132530.4-425914 & 118.7 & 13 25 30.5 & -42 59 14.3 & 496 & 435.8 & 22.3 & $3.68\cdot 10^{37}$ & \\
71 & CXOU J132535.2-430234 & 119.4 & 13 25 35.2 & -43 02 34.1 & 150 & 120.8 & 12.2 & $1.03\cdot 10^{37}$ & \\
72 & CXOU J132517.8-430204 & 120.4 & 13 25 17.9 & -43 02 04.4 & 209 & 181.6 & 14.5 & $2.10\cdot 10^{37}$ & \\
73 & CXOU J132531.9-430302 & 123.1 & 13 25 31.9 & -43 03 02.5 & 56 & 38.2 & 7.5 & $3.34\cdot 10^{36}$ & \\
74 & CXOU J132527.8-425903 & 125.0 & 13 25 27.9 & -42 59 03.9 & 43 & 26.5 & 6.6 & $2.35\cdot 10^{36}$ & \\
75 & CXOU J132528.4-430315 & 126.9 & 13 25 28.4 & -43 03 15.4 & 251 & 211.5 & 15.8 & $1.81\cdot 10^{37}$ & \\
76 & CXOU J132535.5-425935 & 127.6 & 13 25 35.5 & -42 59 35.3 & 609 & 549 & 24.7 & $5.27\cdot 10^{37}$ & GC & pff-gc-214\\
77 & CXOU J132538.3-430205 & 130.4 & 13 25 38.3 & -43 02 05.8 & 2221 & 2158.8 & 47.1 & $1.78\cdot 10^{38}$ & \\
78 & CXOU J132539.4-430058 & 130.6 & 13 25 39.5 & -43 00 58.9 & 39 & 22 & 6.2 & $1.99\cdot 10^{36}$ & \\
79 & CXOU J132517.0-430007 & 131.1 & 13 25 17.1 & -43 00 07.5 & 46 & 33.9 & 6.8 & $3.15\cdot 10^{36}$ & \\
80 & CXOU J132533.3-425913 & 131.5 & 13 25 33.4 & -42 59 13.6 & 101 & 78 & 10.1 & $7.40\cdot 10^{36}$ & \\
81 & CXOU J132515.7-430158 & 139.6 & 13 25 15.7 & -43 01 58.2 & 58 & 44.7 & 7.6 & $7.11\cdot 10^{36}$ & \\
82 & CXOU J132533.6-430313 & 141.2 & 13 25 33.7 & -43 03 13.3 & 188 & 159.3 & 13.7 & $1.36\cdot 10^{37}$ & \\
83 & CXOU J132514.8-430048 & 141.6 & 13 25 14.8 & -43 00 48.7 & 42 & 21.3 & 6.5 & $1.94\cdot 10^{36}$ & \\
84 & CXOU J132540.5-430115 & 142.1 & 13 25 40.6 & -43 01 15.2 & 638 & 587 & 25.3 & $5.02\cdot 10^{37}$ & \\
85 & CXOU J132538.2-430230 & 142.3 & 13 25 38.3 & -43 02 30.5 & 15 & 9.4 & 3.9 & $9.11\cdot 10^{35}$ & \\
86 & CXOU J132523.6-430325 & 143.9 & 13 25 23.6 & -43 03 25.9 & 256 & 221.8 & 16 & $2.38\cdot 10^{37}$ & \\
87 & CXOU J132533.9-425859 & 146.4 & 13 25 34.0 & -42 58 59.9 & 558 & 514.4 & 23.6 & $4.59\cdot 10^{37}$ & GC & pff-gc-159\\
88 & CXOU J132520.0-430310 & 147.1 & 13 25 20.1 & -43 03 10.4 & 367 & 317.4 & 19.2 & $3.95\cdot 10^{37}$ & GC & mrfa-071\\
89 & CXOU J132516.9-425938 & 147.9 & 13 25 16.9 & -42 59 38.8 & 19 & 11.9 & 4.4 & $1.18\cdot 10^{36}$ & \\
90 & CXOU J132532.4-425850 & 148.1 & 13 25 32.4 & -42 58 50.5 & 206 & 180.8 & 14.4 & $1.80\cdot 10^{37}$ & GC & pff-gc-178\\
91 & CXOU J132522.3-425852 & 148.2 & 13 25 22.4 & -42 58 52.2 & 41 & 25.8 & 6.4 & $2.34\cdot 10^{36}$ & \\
92 & CXOU J132541.0-430126 & 148.6 & 13 25 41.1 & -43 01 26.8 & 607 & 574.2 & 24.6 & $4.95\cdot 10^{37}$ & \\
93 & CXOU J132514.0-430121 & 149.6 & 13 25 14.0 & -43 01 21.6 & 63 & 43.4 & 7.9 & $6.44\cdot 10^{36}$ & \\
94 & CXOU J132541.0-430037 & 150.0 & 13 25 41.0 & -43 00 37.7 & 49 & 25.6 & 7 & $2.21\cdot 10^{36}$ & \\
95 & CXOU J132516.8-425932 & 152.7 & 13 25 16.8 & -42 59 32.4 & 34 & 24.3 & 5.8 & $2.29\cdot 10^{36}$ & \\
96 & CXOU J132519.9-430317 & 153.5 & 13 25 19.9 & -43 03 17.2 & 2263 & 2028.1 & 47.6 & $2.10\cdot 10^{38}$ & \\
97 & CXOU J132524.9-430341 & 155.2 & 13 25 24.9 & -43 03 41.2 & 33 & 22.8 & 5.7 & $2.46\cdot 10^{36}$ & \\
98 & CXOU J132527.3-425829 & 159.2 & 13 25 27.3 & -42 58 29.7 & 68 & 48.3 & 8.2 & $4.29\cdot 10^{36}$ & \\
99 & CXOU J132541.9-430142 & 161.0 & 13 25 42.0 & -43 01 42.3 & 32 & 21.7 & 5.7 & $1.98\cdot 10^{36}$ & \\
100 & CXOU J132512.9-430114 & 161.4 & 13 25 12.9 & -43 01 14.7 & 589 & 527.1 & 24.3 & $7.40\cdot 10^{37}$ & GC & mrfa-082\\
101 & CXOU J132520.6-425846 & 162.2 & 13 25 20.6 & -42 58 46.0 & 712 & 676.3 & 26.7 & $6.04\cdot 10^{37}$ & \\
102 & CXOU J132516.4-430255 & 162.5 & 13 25 16.4 & -43 02 55.4 & 290 & 262.1 & 17 & $3.19\cdot 10^{37}$ & \\
103 & CXOU J132538.6-425919 & 162.5 & 13 25 38.6 & -42 59 20.0 & 99 & 70.5 & 10 & $6.21\cdot 10^{36}$ & GC & pff-gc-164\\
104 & CXOU J132512.4-430049 & 167.4 & 13 25 12.5 & -43 00 49.4 & 56 & 39.5 & 7.5 & $5.09\cdot 10^{36}$ & \\
105 & CXOU J132512.0-430044 & 172.7 & 13 25 12.0 & -43 00 44.6 & 343 & 302.3 & 18.5 & $3.73\cdot 10^{37}$ & \\
106 & CXOU J132540.0-430255 & 173.2 & 13 25 40.1 & -43 02 55.5 & 45 & 27.1 & 6.7 & $2.34\cdot 10^{36}$ & \\
107 & CXOU J132527.9-430402 & 173.7 & 13 25 28.0 & -43 04 02.5 & 194 & 167.4 & 13.9 & $1.47\cdot 10^{37}$ & GC & mrfa-050\\
108 & CXOU J132540.4-430251 & 174.4 & 13 25 40.5 & -43 02 51.8 & 38 & 26.7 & 6.2 & $2.37\cdot 10^{36}$ & H$_\alpha$ & mrfa-30\\
109 & CXOU J132540.8-430247 & 175.0 & 13 25 40.8 & -43 02 47.1 & 267 & 234.5 & 16.3 & $1.97\cdot 10^{37}$ & \\
110 & CXOU J132514.0-430243 & 175.8 & 13 25 14.1 & -43 02 43.2 & 141 & 118.9 & 11.9 & $1.29\cdot 10^{37}$ & GC & mrfa-080\\
111 & CXOU J132535.7-430340 & 176.5 & 13 25 35.8 & -43 03 40.9 & 44 & 33.7 & 6.6 & $3.06\cdot 10^{36}$ & \\
112 & CXOU J132512.0-430010 & 180.6 & 13 25 12.0 & -43 00 11.0 & 473 & 441.9 & 21.7 & $3.99\cdot 10^{37}$ & FS & mrfa-85\\
113 & CXOU J132529.4-425809 & 180.6 & 13 25 29.4 & -42 58 09.3 & 28 & 18 & 5.3 & $1.96\cdot 10^{36}$ & GC & pff-gc-155\\
114 & CXOU J132533.8-425821 & 180.8 & 13 25 33.9 & -42 58 21.5 & 78 & 61.6 & 8.8 & $5.72\cdot 10^{36}$ & \\
115 & CXOU J132511.1-430132 & 182.7 & 13 25 11.1 & -43 01 32.3 & 25 & 17.8 & 5 & $2.54\cdot 10^{36}$ & \\
116 & CXOU J132542.7-425943 & 186.6 & 13 25 42.8 & -42 59 43.9 & 28 & 17.3 & 5.3 & $1.60\cdot 10^{36}$ & \\
117 & CXOU J132528.3-430416 & 187.9 & 13 25 28.3 & -43 04 16.5 & 113 & 95.9 & 10.6 & $8.55\cdot 10^{36}$ & \\
118 & CXOU J132511.5-430226 & 192.1 & 13 25 11.6 & -43 02 26.6 & 184 & 161.9 & 13.6 & $1.74\cdot 10^{37}$ & \\
119 & CXOU J132514.5-425858 & 193.9 & 13 25 14.5 & -42 58 58.7 & 22 & 13.7 & 4.7 & $1.37\cdot 10^{36}$ & \\
120 & CXOU J132517.7-430350 & 194.4 & 13 25 17.8 & -43 03 50.6 & 43 & 29.9 & 6.6 & $3.45\cdot 10^{36}$ & \\
121 & CXOU J132545.6-430115 & 197.9 & 13 25 45.7 & -43 01 15.9 & 79 & 43.1 & 8.9 & $3.62\cdot 10^{36}$ & GAL & whh-317\\
122 & CXOU J132524.9-430425 & 199.1 & 13 25 24.9 & -43 04 25.7 & 151 & 132.9 & 12.3 & $1.58\cdot 10^{37}$ & \\
123 & CXOU J132512.1-425918 & 201.8 & 13 25 12.2 & -42 59 18.8 & 34 & 25.3 & 5.8 & $2.57\cdot 10^{36}$ & \\
124 & CXOU J132529.3-425747 & 201.8 & 13 25 29.3 & -42 57 47.8 & 48 & 23.8 & 6.9 & $2.18\cdot 10^{36}$ & GC & whh-22\\
125 & CXOU J132514.8-425840 & 204.1 & 13 25 14.8 & -42 58 40.8 & 33 & 23.2 & 5.7 & $2.23\cdot 10^{36}$ & \\
126 & CXOU J132542.1-430319 & 206.0 & 13 25 42.1 & -43 03 20.0 & 52 & 30.5 & 7.2 & $2.59\cdot 10^{36}$ & GC & mrfa-026\\
127 & CXOU J132538.2-425815 & 208.6 & 13 25 38.2 & -42 58 15.8 & 131 & 107 & 11.4 & $9.14\cdot 10^{36}$ & GC & mrfa-003\\
128 & CXOU J132532.8-430429 & 208.8 & 13 25 32.9 & -43 04 29.4 & 173 & 146.6 & 13.2 & $1.30\cdot 10^{37}$ & \\
129 & CXOU J132546.4-430036 & 209.2 & 13 25 46.5 & -43 00 36.7 & 54 & 31.1 & 7.3 & $2.82\cdot 10^{36}$ & \\
130 & CXOU J132519.0-425759 & 211.5 & 13 25 19.1 & -42 57 59.3 & 55 & 42.9 & 7.4 & $4.28\cdot 10^{36}$ & \\
131 & CXOU J132524.4-425735 & 216.1 & 13 25 24.4 & -42 57 35.6 & 12 & 7.1 & 3.5 & $2.08\cdot 10^{36}$ & \\
132 & CXOU J132513.1-425841 & 216.5 & 13 25 13.2 & -42 58 41.2 & 23 & 17.2 & 4.8 & $1.93\cdot 10^{36}$ & \\
133 & CXOU J13257.82-430059 & 217.3 & 13 25 07.8 & -43 00 59.8 & 24 & 15.9 & 4.9 & $2.06\cdot 10^{36}$ & \\
134 & CXOU J132532.3-430441 & 218.7 & 13 25 32.3 & -43 04 41.3 & 49 & 34.3 & 7.0 & $3.27\cdot 10^{36}$ & \\
135 & CXOU J132507.6-430115 & 219.1 & 13 25 07.7 & -43 01 15.5 & 2028 & 1873 & 45.0 & $1.90\cdot 10^{38}$ & GC & whh-8\\
136 & CXOU J132516.0-430411 & 222.1 & 13 25 16.0 & -43 04 11.1 & 31 & 17.2 & 5.6 & $2.14\cdot 10^{36}$ & \\
137 & CXOU J132547.6-430030 & 223.0 & 13 25 47.7 & -43 00 30.7 & 89 & 45 & 9.4 & $4.18\cdot 10^{36}$ & \\
138 & CXOU J132543.2-425837 & 228.5 & 13 25 43.2 & -42 58 37.6 & 152 & 120.9 & 12.3 & $1.04\cdot 10^{37}$ & GC & pff-gc-062\\
139 & CXOU J132509.3-425917 & 229.0 & 13 25 09.4 & -42 59 17.6 & 78 & 45.4 & 8.8 & $4.13\cdot 10^{36}$ & \\
140 & CXOU J132521.8-430451 & 231.3 & 13 25 21.8 & -43 04 51.2 & 120 & 70.1 & 11 & $7.33\cdot 10^{36}$ & \\
141 & CXOU J132512.0-425830 & 233.2 & 13 25 12.0 & -42 58 30.7 & 63 & 33.6 & 7.9 & $3.02\cdot 10^{36}$ & \\
142 & CXOU J132512.3-425824 & 234.6 & 13 25 12.3 & -42 58 24.5 & 82 & 56.4 & 9.1 & $5.12\cdot 10^{36}$ & \\
143 & CXOU J132547.1-430243 & 234.7 & 13 25 47.2 & -43 02 43.6 & 181 & 151.7 & 13.5 & $1.32\cdot 10^{37}$ & \\
144 & CXOU J132514.8-430418 & 235.4 & 13 25 14.8 & -43 04 18.0 & 89 & 35.8 & 9.4 & $3.89\cdot 10^{36}$ & \\
145 & CXOU J132538.8-430432 & 237.9 & 13 25 38.9 & -43 04 32.1 & 32 & 17.5 & 5.7 & $1.63\cdot 10^{36}$ & \\
146 & CXOU J132522.3-425717 & 238.4 & 13 25 22.4 & -42 57 17.4 & 914 & 845.7 & 30.2 & $7.37\cdot 10^{37}$ & GC & mrfa-208\\
147 & CXOU J132546.3-430310 & 239.1 & 13 25 46.4 & -43 03 10.8 & 406 & 356.8 & 20.1 & $3.35\cdot 10^{37}$ & FS & mrfa-93\\
148 & CXOU J132509.2-425859 & 239.7 & 13 25 09.2 & -42 58 59.5 & 682 & 594 & 26.1 & $5.21\cdot 10^{37}$ & GC & mrfa-215\\
149 & CXOU J132506.3-430221 & 244.4 & 13 25 06.3 & -43 02 21.2 & 974 & 818.5 & 31.2 & $8.26\cdot 10^{37}$ & \\
150 & CXOU J132515.8-425739 & 245.6 & 13 25 15.8 & -42 57 39.9 & 141 & 106.5 & 11.9 & $9.78\cdot 10^{36}$ & \\
151 & CXOU J132505.0-430133 & 248.9 & 13 25 05.0 & -43 01 33.5 & 292 & 239.9 & 17.1 & $2.51\cdot 10^{37}$ & \\
152 & CXOU J132534.2-425709 & 249.9 & 13 25 34.3 & -42 57 09.7 & 47 & 25.2 & 6.9 & $2.46\cdot 10^{36}$ & \\
153 & CXOU J132547.6-425903 & 252.6 & 13 25 47.6 & -42 59 03.8 & 56 & 36.4 & 7.5 & $3.31\cdot 10^{36}$ & \\
154 & CXOU J132529.2-430521 & 253.4 & 13 25 29.2 & -43 05 21.5 & 32 & 19.4 & 5.7 & $3.10\cdot 10^{36}$ & \\
155 & CXOU J132548.5-430258 & 254.6 & 13 25 48.6 & -43 02 58.4 & 33 & 18.8 & 5.7 & $2.02\cdot 10^{36}$ & \\
156 & CXOU J132527.5-430525 & 256.5 & 13 25 27.6 & -43 05 25.3 & 55 & 37.2 & 7.4 & $4.92\cdot 10^{36}$ & \\
157 & CXOU J132538.5-425720 & 258.0 & 13 25 38.6 & -42 57 20.5 & 102 & 57.5 & 10.1 & $5.36\cdot 10^{36}$ & \\
158 & CXOU J132518.8-425708 & 258.7 & 13 25 18.9 & -42 57 08.5 & 91 & 61.5 & 9.5 & $6.02\cdot 10^{36}$ & \\
159 & CXOU J132523.5-425651 & 260.7 & 13 25 23.5 & -42 56 52.0 & 204 & 170.4 & 14.3 & $1.78\cdot 10^{37}$ & \\
160 & CXOU J132504.4-430008 & 261.5 & 13 25 04.4 & -43 00 08.2 & 94 & 69.8 & 9.7 & $7.71\cdot 10^{36}$ & \\
161 & CXOU J132545.5-425815 & 261.7 & 13 25 45.5 & -42 58 15.9 & 242 & 179.5 & 15.6 & $1.73\cdot 10^{37}$ & \\
162 & CXOU J132533.6-430525 & 265.1 & 13 25 33.7 & -43 05 25.4 & 86 & 63.8 & 9.3 & $8.22\cdot 10^{36}$ & FS & HD 116647 fs\\
163 & CXOU J132548.5-430322 & 265.6 & 13 25 48.5 & -43 03 22.8 & 56 & 41.8 & 7.5 & $5.83\cdot 10^{36}$ & \\
164 & CXOU J132539.8-430501 & 268.9 & 13 25 39.8 & -43 05 01.9 & 100 & 86.8 & 10 & $2.20\cdot 10^{37}$ & GC & pff-gc-111\\
165 & CXOU J132547.2-425825 & 270.0 & 13 25 47.2 & -42 58 25.6 & 179 & 139.7 & 13.4 & $1.22\cdot 10^{37}$ & \\
166 & CXOU J132538.0-430513 & 270.4 & 13 25 38.1 & -43 05 13.6 & 98 & 82.2 & 9.9 & $2.00\cdot 10^{37}$ & \\
167 & CXOU J132526.1-425636 & 272.5 & 13 25 26.2 & -42 56 36.7 & 184 & 139.6 & 13.6 & $1.25\cdot 10^{37}$ & \\
168 & CXOU J132549.4-425858 & 273.2 & 13 25 49.5 & -42 58 58.4 & 34 & 16.6 & 5.8 & $1.62\cdot 10^{36}$ & \\
169 & CXOU J132535.2-430529 & 273.3 & 13 25 35.2 & -43 05 29.0 & 18 & 14.2 & 4.2 & $4.67\cdot 10^{36}$ & GC & whh-17\\
170 & CXOU J132527.5-430549 & 281.1 & 13 25 27.6 & -43 05 49.9 & 59 & 35.1 & 7.7 & $4.60\cdot 10^{36}$ & \\
171 & CXOU J132503.5-425928 & 282.5 & 13 25 03.5 & -42 59 28.8 & 29 & 17.2 & 5.4 & $2.36\cdot 10^{36}$ & \\
172 & CXOU J132539.1-425654 & 284.3 & 13 25 39.1 & -42 56 54.0 & 352 & 297.3 & 18.8 & $2.67\cdot 10^{37}$ & \\
173 & CXOU J132507.4-430409 & 285.5 & 13 25 07.5 & -43 04 09.3 & 6878 & 6474.7 & 82.9 & $6.69\cdot 10^{38}$ & FS & Kraft\\
174 & CXOU J132553.5-430134 & 285.7 & 13 25 53.6 & -43 01 34.9 & 73 & 44.4 & 8.5 & $3.95\cdot 10^{36}$ & \\
175 & CXOU J132546.7-425752 & 287.1 & 13 25 46.7 & -42 57 52.6 & 90 & 40.5 & 9.5 & $3.51\cdot 10^{36}$ & \\
176 & CXOU J132502.7-430243 & 289.0 & 13 25 02.7 & -43 02 43.5 & 2412 & 2257.9 & 49.1 & $2.45\cdot 10^{38}$ & \\
177 & CXOU J132532.7-425624 & 290.1 & 13 25 32.8 & -42 56 24.2 & 165 & 115.8 & 12.8 & $1.01\cdot 10^{37}$ & GC & pff-gc-056\\
178 & CXOU J132512.0-425713 & 291.2 & 13 25 12.0 & -42 57 13.3 & 29 & 21.9 & 5.4 & $9.13\cdot 10^{36}$ & \\
179 & CXOU J132555.1-430119 & 301.8 & 13 25 55.1 & -43 01 19.2 & 198 & 153.7 & 14.1 & $1.31\cdot 10^{37}$ & \\
180 & CXOU J132538.4-425630 & 302.2 & 13 25 38.4 & -42 56 30.8 & 150 & 100.3 & 12.2 & $8.94\cdot 10^{36}$ & \\
181 & CXOU J132520.2-425615 & 304.3 & 13 25 20.3 & -42 56 15.4 & 42 & 32.2 & 6.5 & $1.02\cdot 10^{37}$ & \\
182 & CXOU J132552.2-425830 & 312.9 & 13 25 52.2 & -42 58 30.7 & 71 & 21.6 & 8.4 & $1.86\cdot 10^{36}$ & GC & pff-gc-072\\
183 & CXOU J132554.6-425925 & 313.4 & 13 25 54.6 & -42 59 25.8 & 1500 & 1311.5 & 38.7 & $1.10\cdot 10^{38}$ & GC & pff-gc-131\\
184 & CXOU J132549.7-430430 & 315.7 & 13 25 49.8 & -43 04 30.3 & 26 & 18.5 & 5.1 & $5.02\cdot 10^{36}$ & \\
185 & CXOU J132556.8-430044 & 321.8 & 13 25 56.9 & -43 00 44.8 & 319 & 244.3 & 17.9 & $2.07\cdot 10^{37}$ & \\
186 & CXOU J132546.5-425703 & 321.8 & 13 25 46.6 & -42 57 03.4 & 638 & 487.2 & 25.3 & $4.56\cdot 10^{37}$ & GC & pff-gc-168\\
187 & CXOU J132549.1-430447 & 321.9 & 13 25 49.2 & -43 04 47.3 & 60 & 37.2 & 7.7 & $8.91\cdot 10^{36}$ & \\
188 & CXOU J132508.2-430511 & 322.2 & 13 25 08.3 & -43 05 11.3 & 364 & 189.8 & 19.1 & $2.11\cdot 10^{37}$ & \\
189 & CXOU J132509.5-430529 & 327.5 & 13 25 09.6 & -43 05 29.6 & 538 & 418.6 & 23.2 & $4.66\cdot 10^{37}$ & \\
190 & CXOU J132534.4-425549 & 327.6 & 13 25 34.5 & -42 55 49.8 & 239 & 164.9 & 15.5 & $1.48\cdot 10^{37}$ & \\
191 & CXOU J132518.5-425547 & 336.0 & 13 25 18.6 & -42 55 47.8 & 10 & 6.4 & 3.2 & $3.11\cdot 10^{36}$ & \\
192 & CXOU J132553.4-425806 & 336.9 & 13 25 53.4 & -42 58 06.5 & 106 & 40.4 & 10.3 & $3.50\cdot 10^{36}$ & \\
193 & CXOU J132512.8-430606 & 338.8 & 13 25 12.9 & -43 06 06.6 & 43 & 25.1 & 6.6 & $3.07\cdot 10^{36}$ & \\
194 & CXOU J132547.3-425647 & 339.5 & 13 25 47.4 & -42 56 47.4 & 144 & 70.8 & 12 & $6.87\cdot 10^{36}$ & \\
195 & CXOU J132539.4-425546 & 347.6 & 13 25 39.4 & -42 55 46.3 & 94 & 38.9 & 9.7 & $3.44\cdot 10^{36}$ & \\
196 & CXOU J132543.9-430610 & 350.1 & 13 25 43.9 & -43 06 10.0 & 146 & 132.4 & 12.1 & $5.91\cdot 10^{37}$ & \\
197 & CXOU J132507.7-425630 & 353.6 & 13 25 07.7 & -42 56 30.5 & 70 & 56.3 & 8.4 & $2.80\cdot 10^{37}$ & GC & mrfa-216\\
198 & CXOU J132510.1-425608 & 356.5 & 13 25 10.2 & -42 56 08.0 & 39 & 30.2 & 6.2 & $1.61\cdot 10^{37}$ & \\
199 & CXOU J132545.2-425604 & 360.4 & 13 25 45.2 & -42 56 04.5 & 95 & 31 & 9.7 & $2.90\cdot 10^{36}$ & \\
200 & CXOU J132456.1-430258 & 362.3 & 13 24 56.1 & -43 02 59.0 & 144 & 80.9 & 12 & $1.09\cdot 10^{37}$ & \\
201 & CXOU J132557.2-425822 & 364.8 & 13 25 57.2 & -42 58 22.3 & 414 & 291.8 & 20.3 & $3.12\cdot 10^{37}$ & \\
202 & CXOU J132510.6-430624 & 366.3 & 13 25 10.7 & -43 06 24.5 & 398 & 299.5 & 20 & $3.28\cdot 10^{37}$ & \\
203 & CXOU J132522.7-425502 & 370.6 & 13 25 22.7 & -42 55 02.1 & 109 & 92 & 10.4 & $4.22\cdot 10^{37}$ & \\
204 & CXOU J132601.4-430043 & 372.3 & 13 26 01.5 & -43 00 43.5 & 109 & 37.9 & 10.4 & $3.32\cdot 10^{36}$ & \\
205 & CXOU J132554.6-425720 & 374.0 & 13 25 54.6 & -42 57 20.8 & 38 & 16.1 & 6.2 & $2.15\cdot 10^{36}$ & \\
206 & CXOU J132531.4-430720 & 374.2 & 13 25 31.4 & -43 07 20.7 & 33 & 26.6 & 5.7 & $1.24\cdot 10^{37}$ & \\
207 & CXOU J132529.1-425447 & 381.9 & 13 25 29.2 & -42 54 47.3 & 22 & 15.5 & 4.7 & $6.89\cdot 10^{36}$ & \\
208 & CXOU J132545.2-425530 & 389.6 & 13 25 45.2 & -42 55 30.6 & 96 & 36.9 & 9.8 & $3.92\cdot 10^{36}$ & \\
209 & CXOU J132552.6-430545 & 389.6 & 13 25 52.6 & -43 05 45.5 & 45 & 38.1 & 6.7 & $1.79\cdot 10^{37}$ & GC & pff-gc-129\\
210 & CXOU J132503.1-425625 & 390.5 & 13 25 03.1 & -42 56 25.5 & 43 & 29 & 6.6 & $1.43\cdot 10^{37}$ & GC & pff-gc-157\\
211 & CXOU J132557.2-430450 & 393.1 & 13 25 57.2 & -43 04 50.5 & 68 & 44.2 & 8.2 & $1.05\cdot 10^{37}$ & \\
212 & CXOU J132558.6-430430 & 395.6 & 13 25 58.7 & -43 04 30.4 & 1259 & 1180.4 & 35.5 & $2.69\cdot 10^{38}$ & \\
213 & CXOU J132510.0-430655 & 396.1 & 13 25 10.1 & -43 06 55.1 & 137 & 90.2 & 11.7 & $1.22\cdot 10^{37}$ & \\
214 & CXOU J132529.0-430744 & 396.2 & 13 25 29.0 & -43 07 44.7 & 23 & 16.5 & 4.8 & $7.34\cdot 10^{36}$ & \\
215 & CXOU J132549.6-430624 & 397.9 & 13 25 49.7 & -43 06 24.7 & 18 & 10.5 & 4.2 & $4.58\cdot 10^{36}$ & \\
216 & CXOU J132510.2-425510 & 405.4 & 13 25 10.3 & -42 55 10.8 & 80 & 65 & 8.9 & $3.34\cdot 10^{37}$ & \\
217 & CXOU J132513.9-430725 & 405.7 & 13 25 14 & -43 07 25.9 & 222 & 168.2 & 14.9 & $2.72\cdot 10^{37}$ & \\
218 & CXOU J132548.7-425530 & 409.8 & 13 25 48.8 & -42 55 31.0 & 76 & 47.7 & 8.7 & $1.05\cdot 10^{37}$ & \\
219 & CXOU J132557.8-425702 & 413.6 & 13 25 57.9 & -42 57 02.5 & 393 & 296.1 & 19.8 & $4.21\cdot 10^{37}$ & \\
220 & CXOU J132521.2-425413 & 420.9 & 13 25 21.3 & -42 54 13.7 & 113 & 98.8 & 10.6 & $4.82\cdot 10^{37}$ & \\
221 & CXOU J132606.3-430112 & 424.8 & 13 26 06.4 & -43 01 12.3 & 35 & 19.7 & 5.9 & $5.82\cdot 10^{36}$ & \\
222 & CXOU J132545.4-425451 & 424.9 & 13 25 45.4 & -42 54 51.4 & 119 & 80.3 & 10.9 & $1.69\cdot 10^{37}$ & \\
223 & CXOU J132533.2-430810 & 426.3 & 13 25 33.2 & -43 08 10.6 & 38 & 28.1 & 6.2 & $1.25\cdot 10^{37}$ & \\
224 & CXOU J132522.7-430822 & 436.7 & 13 25 22.7 & -43 08 22.2 & 20 & 13.3 & 4.5 & $6.24\cdot 10^{36}$ & \\
225 & CXOU J132501.0-430643 & 443.8 & 13 25 01.1 & -43 06 43.8 & 971 & 853 & 31.2 & $1.16\cdot 10^{38}$ & \\
226 & CXOU J132511.1-430755 & 445.1 & 13 25 11.2 & -43 07 55.9 & 144 & 101.4 & 12 & $1.48\cdot 10^{37}$ & \\
227 & CXOU J132506.8-430736 & 449.7 & 13 25 06.8 & -43 07 36.5 & 86 & 43.1 & 9.3 & $6.10\cdot 10^{36}$ & \\
228 & CXOU J132601.2-430528 & 450.3 & 13 26 01.2 & -43 05 28.0 & 949 & 919.9 & 30.8 & $4.14\cdot 10^{38}$ & \\
229 & CXOU J132544.1-430804 & 453.4 & 13 25 44.1 & -43 08 04.5 & 146 & 135.7 & 12.1 & $6.92\cdot 10^{37}$ & \\
230 & CXOU J132540.4-430820 & 453.7 & 13 25 40.5 & -43 08 20.1 & 18 & 11 & 4.2 & $5.37\cdot 10^{36}$ & \\
231 & CXOU J132550.3-425441 & 460.3 & 13 25 50.3 & -42 54 41.9 & 82 & 34.4 & 9.1 & $4.82\cdot 10^{36}$ & \\
232 & CXOU J132459.0-430648 & 462.6 & 13 24 59.0 & -43 06 48.9 & 74 & 34.9 & 8.6 & $4.96\cdot 10^{36}$ & \\
233 & CXOU J132450.4-430452 & 464.9 & 13 24 50.4 & -43 04 52.3 & 106 & 55.9 & 10.3 & $7.67\cdot 10^{36}$ & \\
234 & CXOU J132557.5-425531 & 470.6 & 13 25 57.5 & -42 55 31.5 & 73 & 38.2 & 8.5 & $5.28\cdot 10^{36}$ & \\
235 & CXOU J132504.0-425431 & 474.2 & 13 25 04 & -42 54 31.5 & 107 & 44.1 & 10.3 & $6.57\cdot 10^{36}$ & \\
236 & CXOU J132609.9-430310 & 479.6 & 13 26 09.9 & -43 03 10.5 & 32 & 17.3 & 5.7 & $4.32\cdot 10^{36}$ & \\
237 & CXOU J132557.5-430659 & 479.6 & 13 25 57.5 & -43 06 59.2 & 23 & 16.7 & 4.8 & $8.07\cdot 10^{36}$ & \\
238 & CXOU J132513.9-425331 & 481.3 & 13 25 13.9 & -42 53 31.5 & 344 & 283.6 & 18.5 & $4.18\cdot 10^{37}$ & \\
239 & CXOU J132548.1-430817 & 483.9 & 13 25 48.1 & -43 08 17.2 & 43 & 29.7 & 6.6 & $1.35\cdot 10^{37}$ & \\
240 & CXOU J132511.3-430843 & 488.6 & 13 25 11.3 & -43 08 43.5 & 51 & 34.2 & 7.1 & $1.11\cdot 10^{37}$ & \\
241 & CXOU J132542.0-425323 & 491.9 & 13 25 42.0 & -42 53 23.1 & 26 & 16.6 & 5.1 & $8.55\cdot 10^{36}$ & \\
242 & CXOU J132510.3-425333 & 493.4 & 13 25 10.4 & -42 53 33.2 & 415 & 334.4 & 20.4 & $4.92\cdot 10^{37}$ & \\
243 & CXOU J132611.8-430242 & 494.0 & 13 26 11.9 & -43 02 42.9 & 280 & 247 & 16.7 & $6.45\cdot 10^{37}$ & \\
244 & CXOU J132535.1-425301 & 494.1 & 13 25 35.2 & -42 53 01.7 & 246 & 233.2 & 15.7 & $1.04\cdot 10^{38}$ & \\
245 & CXOU J132546.7-425340 & 495.2 & 13 25 46.7 & -42 53 40.1 & 209 & 118.5 & 14.5 & $1.64\cdot 10^{37}$ & \\
246 & CXOU J132502.9-425413 & 496.3 & 13 25 02.9 & -42 54 13.2 & 307 & 211.7 & 17.5 & $3.10\cdot 10^{37}$ & \\
247 & CXOU J132605.5-425632 & 499.4 & 13 26 05.5 & -42 56 32.5 & 191 & 132.7 & 13.8 & $1.89\cdot 10^{37}$ & GC & pff-gc-122\\
248 & CXOU J132555.4-430745 & 500.7 & 13 25 55.5 & -43 07 45.8 & 29 & 20.6 & 5.4 & $9.29\cdot 10^{36}$ & \\
249 & CXOU J132614.1-430208 & 513.4 & 13 26 14.1 & -43 02 08.6 & 73 & 55.6 & 8.5 & $2.84\cdot 10^{37}$ & \\
250 & CXOU J132511.0-425257 & 523.7 & 13 25 11.0 & -42 52 57.8 & 50 & 26.5 & 7.1 & $4.44\cdot 10^{36}$ & \\
251 & CXOU J132527.3-430953 & 524.3 & 13 25 27.3 & -43 09 53.1 & 61 & 48 & 7.8 & $2.21\cdot 10^{37}$ & \\
252 & CXOU J132458.9-430831 & 542.7 & 13 24 58.9 & -43 08 31.3 & 167 & 86.8 & 12.9 & $1.33\cdot 10^{37}$ & \\
253 & CXOU J132456.7-430813 & 543.0 & 13 24 56.8 & -43 08 13.7 & 798 & 594.6 & 28.3 & $8.36\cdot 10^{37}$ & \\
254 & CXOU J132539.2-430957 & 543.6 & 13 25 39.2 & -43 09 57.3 & 38 & 24.9 & 6.2 & $1.15\cdot 10^{37}$ & \\
255 & CXOU J132615.9-425846 & 548.4 & 13 26 15.9 & -42 58 46.8 & 530 & 392.4 & 23 & $4.70\cdot 10^{37}$ & \\
256 & CXOU J132546.4-430937 & 549.4 & 13 25 46.5 & -43 09 38.0 & 119 & 102.7 & 10.9 & $4.67\cdot 10^{37}$ & \\
257 & CXOU J132450.4-430722 & 553.0 & 13 24 50.5 & -43 07 22.9 & 271 & 151.4 & 16.5 & $2.17\cdot 10^{37}$ & \\
258 & CXOU J132557.4-425342 & 553.8 & 13 25 57.5 & -42 53 42.2 & 282 & 136.4 & 16.8 & $1.52\cdot 10^{37}$ & FS & GF Blue 1\\
259 & CXOU J132525.9-425152 & 556.4 & 13 25 26.0 & -42 51 52.7 & 160 & 96.2 & 12.7 & $1.44\cdot 10^{37}$ & \\
260 & CXOU J132549.3-425241 & 560.6 & 13 25 49.3 & -42 52 41.4 & 897 & 757.4 & 30 & $1.55\cdot 10^{38}$ & \\
261 & CXOU J132503.1-430924 & 563.6 & 13 25 03.1 & -43 09 24.3 & 163 & 85.2 & 12.8 & $1.28\cdot 10^{37}$ & \\
262 & CXOU J132510.9-425214 & 564.4 & 13 25 11.0 & -42 52 14.8 & 109 & 56.3 & 10.4 & $9.01\cdot 10^{36}$ & \\
263 & CXOU J132534.0-431030 & 565.8 & 13 25 34.0 & -43 10 30.2 & 46 & 34.5 & 6.8 & $1.61\cdot 10^{37}$ & \\
264 & CXOU J132613.0-425632 & 569.9 & 13 26 13.1 & -42 56 32.9 & 142 & 68.8 & 11.9 & $7.77\cdot 10^{36}$ & \\
265 & CXOU J132620.4-425947 & 585.6 & 13 26 20.5 & -42 59 47.2 & 142 & 78.6 & 11.9 & $1.94\cdot 10^{37}$ & \\
266 & CXOU J132619.7-430318 & 586.0 & 13 26 19.7 & -43 03 18.8 & 75 & 52.7 & 8.7 & $2.60\cdot 10^{37}$ & \\
267 & CXOU J132454.4-425326 & 588.5 & 13 24 54.4 & -42 53 26.8 & 65 & 35.2 & 8.1 & $7.91\cdot 10^{36}$ & \\
268 & CXOU J132541.7-425137 & 592.0 & 13 25 41.7 & -42 51 37.3 & 24 & 18 & 4.9 & $8.28\cdot 10^{36}$ & \\
269 & CXOU J132541.9-431041 & 593.6 & 13 25 41.9 & -43 10 41.4 & 642 & 610.1 & 25.3 & $2.86\cdot 10^{38}$ & GC & pff-gc-188\\
270 & CXOU J132544.2-425141 & 595.8 & 13 25 44.3 & -42 51 41.7 & 18 & 12.9 & 4.2 & $5.96\cdot 10^{36}$ & \\
271 & CXOU J132548.4-425156 & 597.4 & 13 25 48.5 & -42 51 56.9 & 25 & 20.8 & 5 & $9.54\cdot 10^{36}$ & \\
272 & CXOU J132531.0-431105 & 597.7 & 13 25 31.0 & -43 11 05.3 & 152 & 139.1 & 12.3 & $7.25\cdot 10^{37}$ & GAL & pff-qso-6\\
\label{tab:list}
\end{longtable}
(1) -- the sequence number; 
(2) -- CXO source name, according to the \textit{CHANDRA}-discovered source naming convention;
(3) -- distance from the center in arcsec; 
(4),(5) -- right ascension and declination, J2000; 
(6) -- total number of counts in the {\tt wavdetect} source cell, source+background;
(7) -- number of source counts after background subtraction
(8) -- statistical error on the number of source counts after background subtraction;
(9) -- X-ray luminosity, 0.5--8 keV, assuming 3.5 kpc distance;
(10) -- source type: GC -- confirmed globular cluster, FS -- foreground star, 
GAL -- background galaxy, H$_\alpha$ -- H$_\alpha$ emmitter; 
(11) -- precise identification and reference: 
pff -- \citet{clean2}, Tables 5 and 9; mrfa -- \citet{clean1}, Tables 1 and 3;
whh -- \citet{WHH} Tables 1 and 2 ; GF -- \citet{clean4} Table 1;
Kraft -- \citet{intro3} Sect. 5.1; HD -- \citet{draper}.\\
Comments: Source \#121 was designated a globular cluster by \citet{clean1}, but
according to \citet{WHH} it is a background galaxy; \citet{clean1} claim to
have removed sources with H$_\alpha$-emission from their list of globular
clusters. However two sources (\#54 and \#108 in our source list) are both
listed as H$_\alpha$-emitters and globular clusters in their tables. We
assume that they are H$_\alpha$ sources and designate them accordingly.
Sources \#146, \#148 and \#197 are included in the list of globular
clusters of \citet{clean1}, although no colours are available. They are
marked as globular cluster in the Table.} 
}


\end{document}